\documentclass[12pt]{article}
\usepackage[utf8]{inputenc}
\usepackage{graphicx}
\usepackage{bm}
\usepackage{epsf}
\usepackage{rotating}
\usepackage{epsfig,graphics,rotate,color}
\usepackage{wrapfig}
\usepackage{amssymb}
\usepackage{amsmath}
\usepackage{amsfonts}
\usepackage{hyperref}
\usepackage[margin=2.5cm]{geometry}

\newcommand{\nua}[1]{\ensuremath{\rlap{\kern-2.5pt\ensuremath{\overset{\scriptscriptstyle(-)}{\phantom{\nu}}}}{\ensuremath{{\nu}_{#1}}}}\xspace}

\begin{document}

\begin{center}
{\bf \Large Sterile Neutrinos: An Introduction to Experiments}\\
~~~~\\
J.M. Conrad$^1$ and M.H. Shaevitz$^2$\\
{\it \small $^1$Massachusetts Institute of Technology, Cambridge, MA}\\
{\it \small $^2$Columbia University, New York, NY}
\end{center}

\vspace{0.1in}

\noindent {\bf Abstract:}   {\it This paper is written as one chapter in a collection of essays on neutrino physics for beginning graduate students.   The text presents important experimental methods and issues for those interested in searches for sterile neutrinos.   Other essays in the collection, written by other authors, will cover introduction to neutrinos in the Standard Model, a description of the theory, and discussion of details of detectors, thus these aspects are not covered here. However, beyond these points, this represents a self-contained tutorial on experimental studies of sterile neutrino oscillations, covering such issues as signals vs. limits,  designing experiments, and performing and interpreting global fits to the oscillation data.}

\vspace{0.2in}

\tableofcontents

\newpage

\section{Theoretical Models with Sterile Neutrinos}

Neutrinos are unique in the standard model (SM) in that they interact solely through the weak interaction.  As a consequence, in the SM, only the left-handed neutrino is active and part of a weak isospin doublet with its partner charged lepton.  A question then arises as to whether right-handed neutrinos exist and if they do, how do they fit into the SM?  In the SM, right-handed neutrinos, if they exist, would be weak isospin singlets with no weak interactions except through mixing with the left-handed neutrinos.  For this reason, the right-handed neutrinos are referred to as ``sterile'' neutrinos.  

In fact, the sterile neutrino is more broadly defined as a neutral lepton with no ordinary weak interactions except those induced by mixing. So, the right-handed neutrino is a prime example, although not the only case. Note, however, that sterile neutrinos can possibly interact through Yukawa couplings with the Higgs boson or through other new physics interactions beyond the SM.   So it is not necessarily true that sterile neutrinos have no interactions at all.

Explaining in detail how sterile neutrinos arise in theories is beyond the scope of this chapter.  Here,  we briefly review the basic theoretical ideas before embarking on the experimental questions related to sterile neutrino searches, which is the primary goal of this text.   Thus, this section is a bit technical, and students who are not already familiar with the theory of neutrino mass may wish to skip onward to the experimental discussion that follows.   A useful tutorial on introducing neutrino mass appears in Ref.~\cite{wikineutrinomass}.  For those who want to learn more about the theory than is discussed here, a good detailed review appears in Ref.~\cite{whitepaper}.     

In the SM, neutrinos are massless, since a Dirac mass term would require the existence of right-handed neutrinos.  On the other hand, neutrino oscillations, which have been unambiguously observed \cite{pdg}, require that the active neutrinos must have mass.  Generating a non-zero neutrino mass can be accomplished by adding a mass-term to the SM Lagrangian.  
A Dirac mass-term similar to the charged fermion terms requires the existence of right-handed neutrinos.  Then, a term with a right-handed neutrino combined with a left-handed neutrino and a SM Higgs field would produce a Dirac mass.  To produce such a Dirac mass for the neutrino below the current upper limit from tritium decay studies of 2 eV \cite{pdg} would require a Yukawa coupling of order $10^{-12}$, which would be very much smaller than of the other couplings.  Explanation of these small couplings has led to theoretical models where the neutrino mass terms involve higher dimensional operators, non-perturbative effects or warped extra dimensions \cite{whitepaper}.

If neutrinos are Majorana particles, where the neutrino and antineutrino are the same particle species, then a Majorana mass terms is also possible.  Such a Majorana term would violate lepton number by two units bringing in the possibility of neutrino to antineutrino transitions.  A Majorana term for a left-handed neutrino couples the left-handed neutrino with its conjugate and a Higgs triplet state, since such a transition would violate weak isospin by one unit.  Instead of coupling to a Higgs triplet state, a Majorana term could also be constructed of  higher-dimensioned terms involving a coupling of the left-handed neutrino to two Higgs doublets. Finally, the right-handed neutrino can also have a Majorana mass term where the neutrino couples with its conjugate since, as a singlet, this term does not violate weak isospin. 

The combination of Dirac and Majorana mass terms for neutrinos can lead to an interesting scenario where the diagonalization of the mass matrix leads to active neutrinos with small masses of $(m_{Dirac})^2/M_{heavy}$ and also a heavier sterile neutrinos that mix with the active neutrinos.  This is referred to as the See-Saw Mechanism and may be the explanation for why neutrino masses are so small relative to the other fermions that have masses on the order of $m_{Dirac}$. On the other hand, the sterile neutrinos that are most naturally suggested by the See-Saw Mechanism are very heavy compared to all known particles.  Nevertheless, light sterile neutrinos are a possibility within many models using the See-Saw framework \cite{whitepaper}. 

These theoretical speculations about the existence of sterile neutrinos and how they might impact neutrino masses and mixing are key areas currently being investigated in particle physics.  The experimental discovery of sterile neutrinos and their properties would have a major impact on these speculations and would likely lead to insights as to how to add them to the standard model.  In addition, sterile neutrinos can have a major impact on interpreting other neutrino measurements such as the search for CP violation in long baseline experiments \cite{boris} and modeling of neutrino production in core-collapse supernovae \cite{Warren}.

\section{A Brief Tutorial on Signals}

Before discussing sterile neutrino oscillation signals, it is useful to review how oscillation results are presented.   We will consider this within a generic oscillation model between two flavors of neutrinos $\alpha$ and $\beta$. Then we will expand on this to discuss signals with three active flavors and one or more sterile neutrinos.

\subsection{Reviewing the Basics of the Two-neutrino Oscillation Example \label{basics}}

In a two neutrino oscillation model, the flavor eigenstates ($\alpha /\beta$ subscripts) can be written as a function of the mass eigenstates (1/2 subscripts) as: 

\[
\begin{array}{l}
\nu _\alpha=\cos \theta \;\nu _1+\sin \theta \;\nu _2 \\ 
\nu _\beta =-\sin \theta \;\nu _1+\cos \theta \;\nu _2
\end{array}
\]
where $\theta$ is the ``mixing angle.''   The implication of mixing is that a pure flavor (weak) eigenstate born through a weak decay can
oscillate into the other flavor as the state propagates in space. This
oscillation is due to the fact that each of the mass eigenstate components
propagates with a different frequency, assuming the masses are different, $\Delta
m^2=\left| m_2^2-m_1^2\right|>0$.  Using quantum mechanics, one can calculate the two-neutrino oscillation probability for $\nu_\alpha \rightarrow \nu_\beta$ as:
\begin{equation}
{\rm Prob}\left( \nu _\mu \rightarrow \nu _e\right) = \sin ^22\theta \;\sin
^2\left( \frac{1.27\;\Delta m^2\left( {\rm eV}^2\right) \,L\left({\rm km}%
\right) }{E \left({\rm GeV}\right) }\right). 
  \label{prob}
\end{equation}
In this equation,  
$L$ is the distance from the source, and $E$ is the neutrino energy. 
Examining this equation, one can see that the oscillation wavelength will
depend upon $L$, $E$, and $\Delta m^2$.  The amplitude will depend
upon $\sin^2 2\theta$.

Looking at Eq. ~\ref{prob}, one can see that the designer of an experiment would like to arrange for $1.27 \Delta m^2 L/E = \pi/2$ because this maximizes the oscillation probability. The $L/E$ value that maximizes this probability, $L/E = \pi/(2.54 \Delta m^2)$, is called the ``oscillation maximum'' value.  Of course, the probability function has a number of maxima but the first one usually dominates and others can be partially washed out due to energy and position smearing.  

However, when one is searching for new physics, one does not necessarily know the exact $\Delta m^2$ of interest.  Thus there is some art associated choosing the detector location and the beam energy.  The experiment must be designed 
with a relatively large value of the ratio $L/E$ in order to enhance the $\sin^2 (1.27 \Delta m^2 L/E)$ term.
However if $L/E$ is too large in comparison to $\Delta m^2$, then  
oscillations occur rapidly.  Because experiments have finite
resolution on $L$ and $E$, and a spread in beam energies, the 
$\sin^2 (1.27 \Delta m^2 L/E)$ term averages to $1/2$ when $\Delta m^2 \gg
L/E$ and one loses
sensitivity to $\Delta m^2$.    
Lastly, since the oscillation probability is
directly proportional to $\sin^2 2\theta$, if the mixing angle is
likely to be small, then a design that results in high statistics is required to observe the tiny oscillation
signal.   

There are two types of oscillation searches: ``disappearance'' and
``appearance.''  
Consider a pure source of neutrinos
of type $\alpha$. In a disappearance experiment, one looks for a deficit in the expected flux of $\nu_\alpha$ at a detector located downstream from the source, hence later in time.   Another way to say this, is ``if it started as a $\nu_\alpha$, did the neutrino stay a $\nu_\alpha$?'' and thus, it is denoted as $\nu_\alpha \rightarrow \nu_\alpha$.
On the other hand, appearance experiments search for one flavor turning into the other flavor, $\nu_\alpha
\rightarrow \nu_\beta$, by directly observing interactions of neutrinos of
type $\beta$. In either case,  the signal for oscillations will be most persuasive if the deficit
or excess has the ($L/E$) dependence predicted by the neutrino
oscillation formula (equation \ref{prob}). 

Now consider how the results of neutrino oscillations can be presented.    In this two-neutrino model, which was the first model for oscillations that was developed, there are two theoretical parameters ($\Delta m^2$ and the mixing angle) and so the results are traditionally shown on the plane of $\Delta m^2$ {\it vs.} $\sin^2 2\theta$.      
Let's say that a hypothetical, perfect (no systematic error) two-neutrino oscillation experiment sees no
oscillation signal, based on $N$ expected events without oscillations.
The experimenters can rule out the probability for oscillations (either appearance or disappearance) within some error or
confidence level (CL).   A typical choice of confidence level is 90\% CL, so
in this case of statistical-uncertainty-only, the approximate limiting probability is $P < 1.28 \sqrt{N}/N$.  The 1.28 factor corresponds to a one-sided Gaussian probability for the expected $N$ value to have an upward fluctuation at the 10\% probability level.  More rigorous methods for setting CL regions will be described later with respect to global fitting techniques.   There
is only one measurement and there are two unknowns, so
this translates to an excluded region within $\Delta m^2$ -- $\sin^2 2\theta$
space.  This is indicated by a
solid line, with the excluded region on the right.   At high $\Delta
m^2$, the limit on $\sin^2 2\theta$ is given by twice the above $P$ value, since $\langle \sin^2 (1.27 \Delta m^2 L/E) \rangle$ averages to $1/2$.    The $L$ and $E$ of the experiment drive the low
$\Delta m^2$ limit, and this only improves by the fourth root of the statistics associated with $N$.   Thus to improve reach in the mixing angle, the designer must increase the total number of events,  while to improve the reach in low $\Delta m^2$, the designer should focus on adjusting the $L/E$ ratio.

If an appearance experiment measures an excess of $\nu_\beta$ events, $N_{excess}$, with some uncertainty, $\delta N_{excess}$, in a $\nu_\alpha \rightarrow \nu_\beta$ search, then the experiment can claim a signal if the excess is greater than its uncertainty.  (This is also true for a disappearance signal where a significant deficit in events, $N_{deficit}$, observed.)  This observed signal can be converted into an average oscillation probability using the number of fully transmuted  $\nu_\alpha$ events, $N_{full trans}$.  $N_{full trans}$ is the number of $\nu_\beta$ events that would be observed in the experiment if all of the $\nu_\alpha$ neutrinos changed into $\nu_\beta$ neutrinos.  To calculate $N_{full trans}$, one has to take the flux of $\nu_\alpha$ neutrinos into the experiment and have them interact with the $\nu_\beta$ cross section and detection efficiency.  The average oscillation probability is then given by $P\pm \delta P = N_{excess}/N_{full trans} \pm \delta N_{excess}/N_{full trans}$.  This type of average oscillation probability is often given as a measure of an experiment's signal with a significance of $N\sigma$ where $N = P/ \delta P$.

In reality,  imperfections of an experiment affect the limits which can 
be set.
Background sources may introduce $\nu_\beta$ into the $\nu_\alpha$ beam.
Misidentification of the interacting neutrino flavor
in the detector may mimic oscillation signatures. In addition, systematic
uncertainties in the relative acceptance versus distance and energy must
be understood and included in the analysis of the data.  These
systematics are included in the 90\% CL excluded regions presented by
the experiments.

Excluded and allowed regions are determined using a test statistic that depends on the likelihood for obtaining a given data set depending on the assumed oscillation parameters.  One then determines the allowed oscillation parameter space where the likelihood is greater than the desired CL.  Since the oscillation probability in Eq. \ref{prob} is non-linear in the oscillation parameters, determining the critical value for the test statistic at a given CL usually requires assumptions and studies of simulated data samples.  

A common example of a test statistic for a $\Delta m^2$ {\it vs.} $\sin^22\theta$ point is the $\Delta \chi^2 = \chi^2_{\Delta m^2 / \sin^22\theta} - \chi^2_{BestFit}$.  This statistic can be shown in a simple Gaussian error case to follow a $\chi^2$ distribution with the degrees of freedom (dof) equal to the number of fit parameters, which for this case would be 2 corresponding to $\Delta m^2$ and $\sin^22\theta$.  In reality due to the non-linear nature of the oscillation probability, the limiting value of $\Delta \chi^2$ for a given CL needs to be determined using simulated data sets.  For example, at high $\Delta m^2$, the value of $\Delta m^2$ does not affect the oscillation probability and, in this region, the $\Delta \chi^2$ test statistic will follow a $\chi^2$ distribution with 1 dof.

The  ``sensitivity'' of an experiment is
defined as the average expected limit if the experiment were performed
many times with no true signal (only background).    This leads to an important complication.
When the actual background is low compared to prediction there can be a significant difference between the limit and the experimental sensitivity.  In some cases, this is due to statistical fluctuation of the background.  In this case, the experiment just ``got lucky'' and the limit represents the true confidence level.   On the other hand, if the true background is due to a systematic effect that leads to an over-prediction, then the extracted limit is not representative of the true space explored.    
To address this concern, in an influential paper on statistics for oscillation experiments,  Feldman and Cousins \cite{Feldman_Cousins} suggested that 
when an experiment sets a significantly better limit than the
sensitivity, the experiment should also indicate the sensitivity on
the plot. This allows readers to draw conclusions based upon their own
opinion of what is acceptable.    An improved approach was recently introduced by the LHC experiments for their electroweak and Higgs searches.  A band is defined around the line indicating the sensitivity that encompasses a given percentage, for example 90\%, of the simulations with no true signal.   One would then expect the result of an experiment to largely lie within this band, at 90\% CL.   The result may cross from one side to the other side of the band, and may even depart from the band, for some best fit parameters.  But overall, the result should be consistent with the band. A result that lies significantly outside of the band, beyond the expectation for the stated confidence level of the band, indicates either a problematic null result or a signal.

We emphasize that limits are not ``hard cut-offs.''    There is some probability that there is a solution outside of the line defined by the limit.    However, that signal will have lower confidence level than the confidence level of the cut-off.   This will be an important point below, when we compare results from multiple experiments.

\subsection{Two Examples of Results Interpreted Within the Two Neutrino Model \label{2examples}}

\begin{figure}[t]
\begin{center}
{\includegraphics[width=0.95\textwidth]{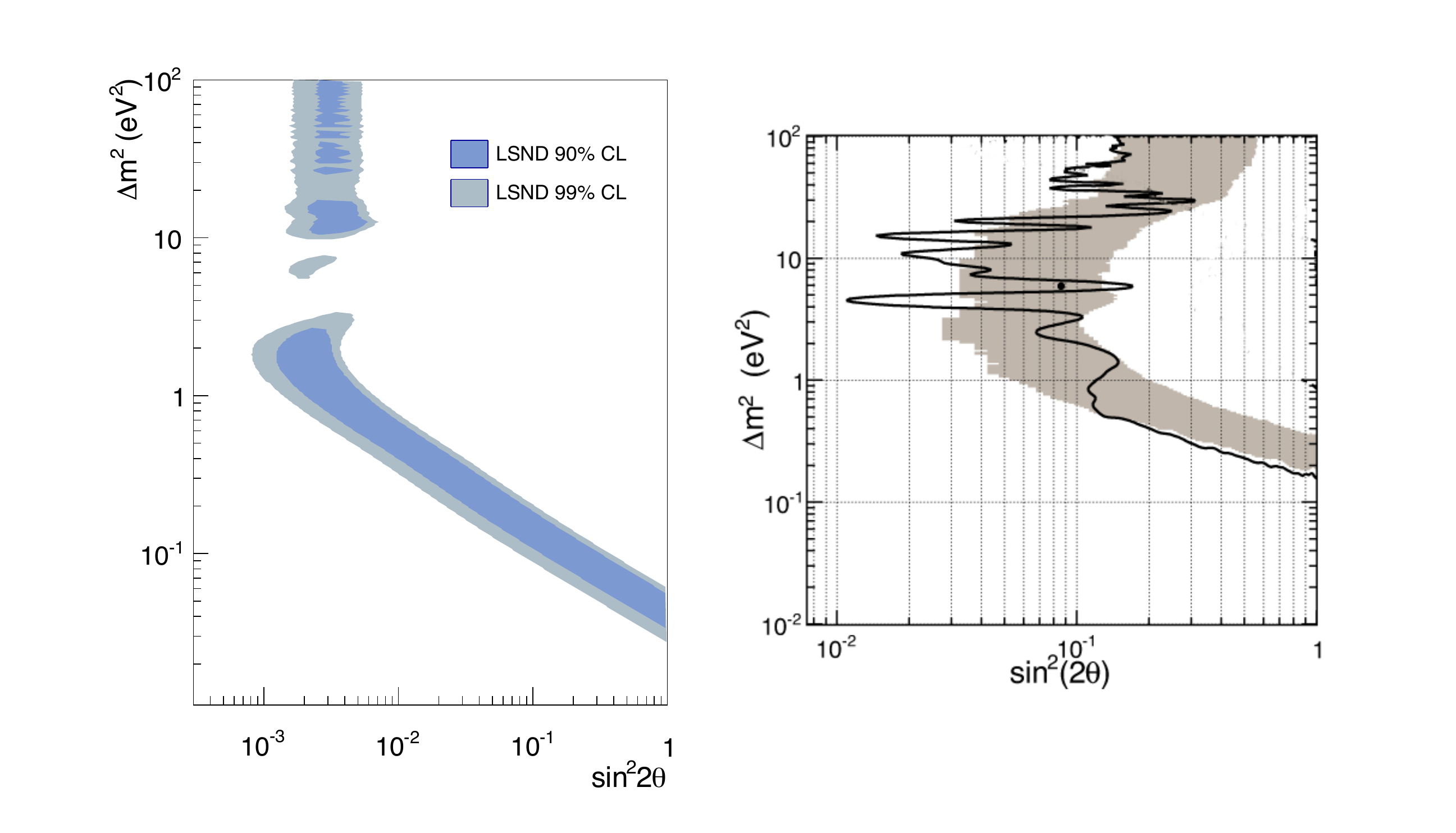}}
\end{center}
\caption{\it Two examples of results from oscillation experiments.   Left:  The LSND allowed region for $\bar \nu_\mu \rightarrow \bar \nu_e$ appearance.   The shaded regions indicate 90\% (blue) and 99\% (grey) CL allowed regions \cite{LSNDprimaryresult}.    Right:  The SciBooNE/MiniBooNE 90\% CL limit for $\bar \nu_\mu \rightarrow \bar \nu_\mu$ disappearance \cite{MBSBantinu}.  The region to the right of the solid line is excluded.    The shaded area indicates the region where 68\% of experiments with no observed signal are predicted to lie, and the sensitivity of the experiment is defined to be the center of the shaded band.
\label{fig:LSNDMBSB}}
\end{figure}

The above discussion was fairly abstract, so let's now consider two real cases of results from oscillation experiments.   The first is an experiment searching for a $\bar \nu_\mu \rightarrow \bar \nu_e$ appearance signal,  
LSND (the Liquid Scintillator Neutrino Detector) \cite{LSNDprimaryresult}.    The second is a pair of detectors,  SciBooNE and MiniBooNE, used to search for a disappearance signal of muon antineutrinos,  $\bar \nu_\mu \rightarrow \bar \nu_\mu$ \cite{MBSBantinu}.    The two results are shown in Fig.~\ref{fig:LSNDMBSB}, and are among a large set of experiments we will discuss.

\begin{figure}[t]
\begin{center}
{\includegraphics[width=0.75\textwidth]{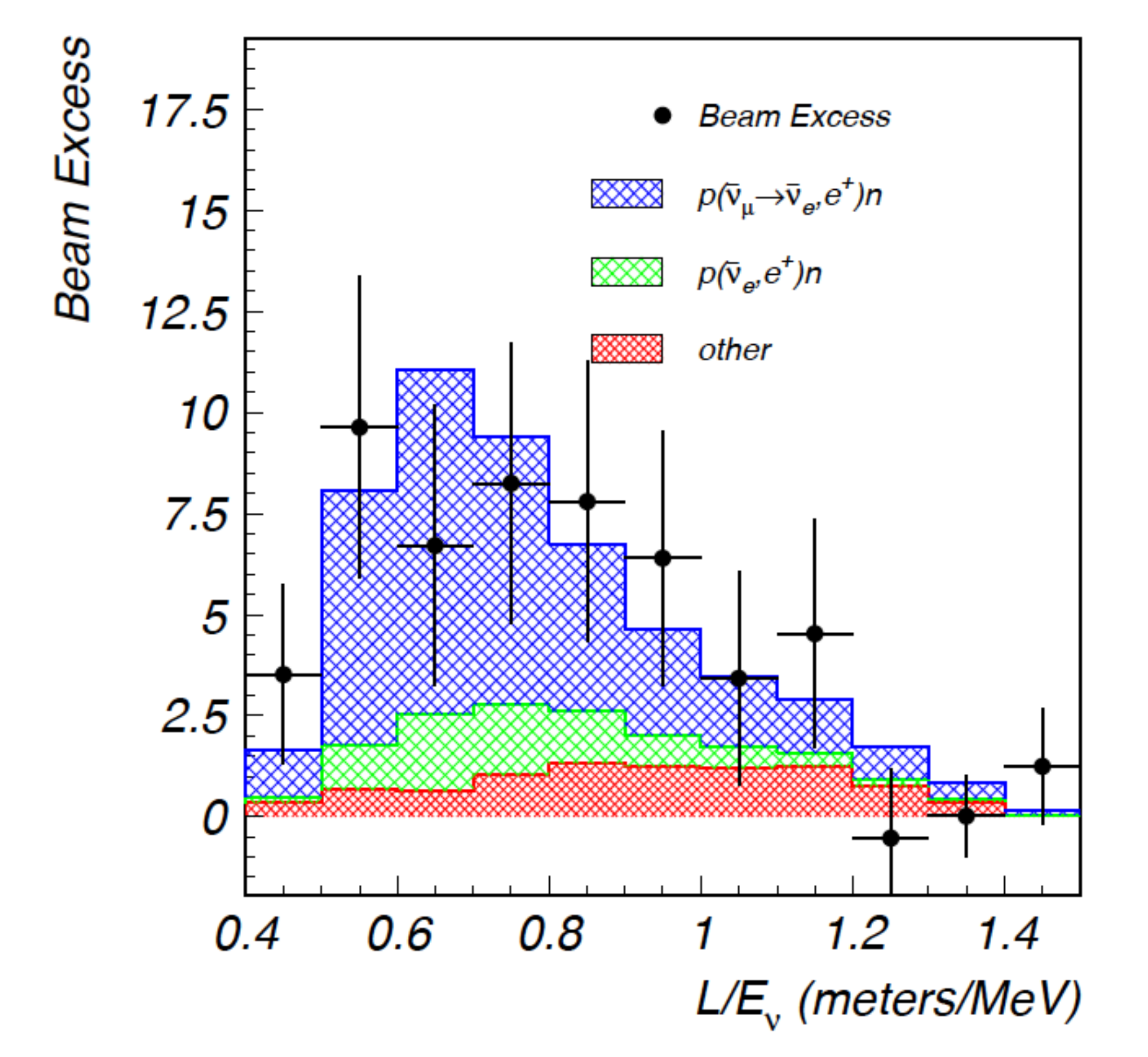}}
\end{center}
\caption{\it The LSND excess \cite{LSNDprimaryresult}, indicating possible $\bar \nu_\mu \rightarrow \bar \nu_e$ appearance.   The data points have statistical and systematic error added in quadrature.   The signal is not background subtracted.  The red and green histograms indicate the background from intrinsic $\bar \nu_e$ in the beam and from other sources, respectively.    The blue histogram is an example of a potential  $\bar \nu_\mu \rightarrow \bar \nu_e$ signal that, when added to the predicted backgrounds, is representative of the data.    This potential signal corresponds to one point in the allowed region in $\Delta m^2$ and $\sin^2 2\theta$.
\label{LSNDsignal}}
\end{figure}

The LSND $\bar \nu_\mu \rightarrow \bar \nu_e$ appearance result was the first of a set of unexpected signals observed at $\Delta m^2 \sim 1$ eV$^2$, and has a significance of 3.8$\sigma$.    This experiment ran at the Los Alamos LAMPF accelerator 
National Laboratory from 1993 and 1998.  The decay-at-rest (DAR)
beam was produced by impinging 800 MeV protons on a beam dump, resulting in 
$\pi^+$s which stop and decay to produce $\mu^+$s, which, in turn, 
stop and decay to produce $\bar \nu_\mu$ and $\nu_e$.   Because the $\pi^-$s produced in the dump
capture, the
beam has a $<8\times10^{-4}$ contamination of $\bar \nu_e$.  
The resulting neutrinos were observed in a detector located 30 m downstream of
the beam dump. Events with neutrino energies between 20 and 52 MeV were used in the analysis.  
In the detector, 1220 phototubes surrounded a
cylindrical detector that was filled with 167 tons of mineral oil, lightly
doped with scintillator.  The signature of $\bar \nu_e$ appearance
was $\bar \nu_e + p \rightarrow e^+ + n$, a process called ``inverse beta decay."  This resulted in a
two-component signature in the detector: 1) the initial Cherenkov and scintillation light
associated with the $e^+$ and 2) the later scintillation light
from the $n$ capture on hydrogen which produces a 2.2 MeV $\gamma$.  The
experiment observed $87.9 \pm 22.4 \pm 6.0$ events
\cite{LSNDprimaryresult} above expectation.    A plot of the data, shown with a stacked plot of background (green and red) and fitted signal (blue),  is shown in Fig.~\ref{LSNDsignal}.

For LSND, the ratio of the distance to the detector to the beam energy,  $L/E$, is approximately 0.8 m/MeV.  Thus,
from the two-generation oscillation formula, Eq.~\ref{prob}, one can
see that this experiment is approximately sensitive to $\Delta m^2 \ge 0.1$ eV$^2$ (this is also seen in Fig.~\ref{fig:LSNDMBSB}).  The oscillation probability measured by LSND for the excess was $(2.64 \pm 0.67 \pm 0.45)\times 10^{-3}$. The resulting allowed regions are shown in Fig.~\ref{fig:LSNDMBSB}, left.  This plot shows ``allowed regions,'' which are  fully enclosed contours, at 90\% and 99\% CL.    Below about 1 eV$^2$ one can see an extended contour that has the typical slope of -0.5 on the log-log plot corresponding to the region where the oscillatory sine factor that depends on $\Delta m^2$ in Eq. \ref{prob} has a small argument.   This is the range where the experiment is sensitive to the small oscillation probabilities before the first oscillation maximum.  The small $\Delta m^2$ limit at $\sin^22\theta = 1.0$ is given by 
\begin{equation}
\Delta m^2 = \frac{\sqrt{Prob}}{1.27L/E},
\end{equation}
which for the above LSND probability gives 0.05 eV$^2$.   Above 1 eV$^2$, one sees that the LSND result forms islands and exclusion regions because of the oscillatory behavior associated with Eq. \ref{prob}. At high $\Delta m^2$, one reaches the region of rapid oscillations where the oscillatory factor averages to a 0.5.  The pattern of this LSND allowed region is typical for a single detector experiment combined with a neutrino source that has a small energy spread.    

While LSND was a single detector experiment with an allowed region,   the SciBooNE/MiniBooNE $\bar \nu_\mu \rightarrow \bar \nu_\mu$ shown in Fig.~\ref{fig:LSNDMBSB}, right, shows typical features of a two-detector experiment where the result is an exclusion.    The two-detector design is often used in cases where the flux is rather poorly predicted, which is often the case for decay-in-flight (DIF) beams.   To produce DIF beams, accelerated proton are directed onto a target where pions and kaons are produced.  The charged pions and kaons are focused magnetically into a gas- or vacuum-filled decay pipe where the mesons decay into muons and muon neutrinos.  Generally, this type of beam is mainly composed of muon neutrinos with a small contamination (0.5\% to few a few percent) of electron neutrinos from muon and kaon decay.

The SciBooNE experiment has a 10.6 ton, scintillator-strip neutrino detector located 100 m from the primary proton target that produced the beam. The fine segmentation allowed the position of the neutrino interaction vertex
to be well identified.    Located at 540 m from the proton target was a second detector,the 800 ton MiniBooNE mineral-oil based neutrino detector.  In the case of MiniBooNE, events were detected using mainly Cherenkov light associated with the outgoing particles, since the oil was undoped with scintillator.  The Cherenkov light formed rings on the 1280 photomultiplier tubes around the periphery of the spherical detector.  The amount of light and Cherenkov ring geometry were used to measure the particle energies and direction as well as the particle type: muon, pion, or electron.  The combination of MiniBooNE with SciBooNE allows a much more precise search for neutrino disappearance than MiniBooNE could perform independently, since the neutrino flux and energy distribution can be measured in the near SciBooNE detector before oscillations occur.  Many systematic uncertainties associated with the neutrino flux and interaction cross sections are reduced or eliminated by comparing a near and far detector.  Detector effects are most reduced if the two detectors use the identical detection technology; however this was not the case for the SciBooNE/MiniBooNE experiment.  For practical reasons, it is often true that the near and far detectors are constructed with different technologies.

Fig. \ref{fig:LSNDMBSB}, right, shows the results for the SciBooNE/MiniBooNE $\bar \nu_\mu$ disappearance search.  The near and far detectors saw consistent results,  so no sign of oscillations was detected.  The experimental result is, therefore, a limit at 90\% CL as shown in the figure. Oscillations are excluded in the region to the right of the black line.   This differs from the LSND example in that there is no fully enclosed space within the parameters, which is to say that the null result, which is off the plot to the left, is allowed.  The band in this figure is indicating the expectation if one runs many experiments, as discussed above. This band encompasses 90\% of the limits obtained from a collection of simulated data runs.   The average of these experimental results, along the center of the band, is the sensitivity (not explicitly shown in this plot). The obtained limit is consistent with the estimated sensitivity and shows classic expected behavior:  most of the limit is contained within the band, with several crossings of the sensitivity, depending on the parameter values.   Within the statistics, one expects the limit to exit the band for a few parameter values, and this is observed.   

For $\Delta m^2 < 1$ eV$^2$, the MiniBooNE/SciBooNE limit curve shows the same characteristics as seen in the LSND result.  For this measurement, the disappearance probability limit is at about the 5\% level, which would predict a $\Delta m^2 = 0.22$ eV$^2$ at $\sin^22\theta = 1.0$.  Above $\Delta m^2 = 1$ eV$^2$, the limit shows a complicated curve associated with the oscillatory behavior of Eq. \ref{prob} for detectors at two different L values and with uncorrelated statistical fluctuations.  At high $\Delta m^2$, in this case above 20 eV$^2$, there are rapid oscillations in both the near and far detector and the sensitivity grows worse with higher $\Delta m^2$ since the two detector comparison no longer helps.  On the other hand, limits can still be set in this region since the absolute neutrino flux is known to some accuracy from the beam and cross section modeling, it is just that the limit is less stringent due to the larger systematic error.

\subsection{Why Go Beyond a Three Neutrino Model?}

\begin{figure}[t]
\begin{center}
{\includegraphics[width=0.5\textwidth]{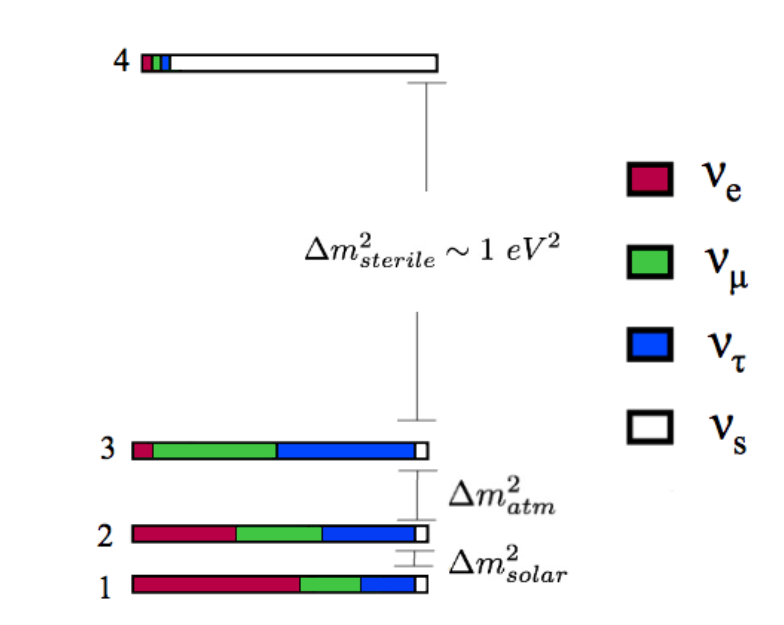}}
\end{center}
\caption{\it Commonly used illustration of four mass states and their relationship to four flavor components.  
\label{nustack}}
\end{figure}

The LSND result shown in Fig.~\ref{fig:LSNDMBSB}, left, is consistent with oscillations with $\Delta m^2 \gtrsim 0.1$ eV$^2$.  This is substantially larger than the $\Delta m^2$ values that have been measured for the ``solar'' and ``atmospheric'' oscillations:  $\Delta m^2_{solar}=7.5 \times 10^{-5} {\rm eV}^2$ and $\Delta m^2_{amtmos}=2.3 \times 10^{-3} {\rm eV}^2$ \cite{nufit}.   These latter are two well-established signals, with significance much higher than 5$\sigma$, observed in many experiments, and celebrated with the
2015 Nobel Prize in physics.   As we discuss in this section,  the solar and atmospheric oscillations are consistent with a model of three neutrinos.     We show here that a three neutrino model cannot explain a third $\Delta m^2$ value.    As a result, LSND, and other experiment that show oscillations with $\Delta m^2 \sim 1$ eV$^2$ are said to have ``anomalous results."

Let's begin by considering only the solar an atmospheric oscillation results.
A model that incorporates two distinct solar and atmospheric $\Delta m^2$ is consistent with three neutrinos, with mixing between the mass and flavor eigenstates given by:
\[
\left( 
\begin{array}{l}
\nu _e \\ 
\nu _\mu \\ 
\nu _\tau
\end{array}
\right) =\left( 
\begin{array}{lll}
U_{e1} & U_{e2} & U_{e3} \\ 
U_{\mu 1} & U_{\mu 2} & U_{\mu 3} \\ 
U_{\tau 1} & U_{\tau 2} & U_{\tau 3}
\end{array}
\right) \left( 
\begin{array}{l}
\nu _1 \\ 
\nu _2 \\ 
\nu _3
\end{array}
\right). 
\]
The oscillation probability is then:
\begin{eqnarray}
{\rm Prob}\left( \nu _\alpha \rightarrow \nu _\beta \right)&
\hspace{-2.in}=\delta_{\alpha \beta }- \nonumber \\
& 4\sum\limits_{j>\,i}U_{\alpha \,i}U_{\beta \,i}U^*_{\alpha
\,\,j}U^*_{\beta \,\,j}\sin ^2\left( \frac{1.27\;\Delta m_{i\,j}^2 \,L }{E }%
\right)  \label{3-gen osc}
\end{eqnarray}
where $\Delta m_{i\,j}^2=\left| m_i^2-m_j^2\right| $ .
Note that there are three different 
$\Delta m^2$ values, but only two are independent.   In this model,  $\Delta m^2_{12}+\Delta m^2_{23}=\Delta m^2_{13}$.   Thus the LSND result cannot be introduced into this model. 

This was a slightly simplistic argument.   The more general case allows the atmospheric result to be due to 
a mixture of high (LSND-range)  and low  (solar-range) $\Delta m^2$ values.    But global fits have shown that this model disagrees with several data sets \cite{sorel}.  So this is ruled out.

\subsection{Introducing a Sterile Neutrino: the 3+1 Models}

As a result, one is forced to introduce new physics if one accepts that LSND, and the other consistent, but less significant, anomalies that we will discuss in Sec.~\ref{currentresults}, are due to oscillations.    The simplest next step is to introduce one additional neutrino mass state, connected to one additional flavor state through the unitary mixing matrix:
\begin{equation}
U_{3+1} = \begin{bmatrix}
U_{e1} & U_{e2} & U_{e3} & U_{e4} \\ 
\vdots & & \vdots & U_{\mu4} \\
\vdots & & \vdots & U_{\tau4} \\
U_{s1} & U_{s2} & U_{s3} & U_{s4}
\end{bmatrix}, \label{4mixmx}
\end{equation}
where the new flavor is denoted with ``s'' in the matrix.   
This explicitly allows the new flavor to mix with the three known neutrino flavors, and thereby be produced in neutrino oscillations.   Fig.~\ref{nustack}, left, shows a common illustration of the four mass states (bars) divided up into their four flavor components (colors).   Because there are three states that are relatively similar in mass in this model and one mass state that is larger, this is called a 3+1 model.

Introducing a new neutrino flavor into the theory has important consequences outside of oscillation studies. First, the precise measurements of the $Z^0$ width \cite{pdg} made at LEP and
SLD, determined that there are only three families of
light-mass, weakly-interacting neutrinos.  Hence, the additional neutrino, if it is lighter than half the mass of the $Z$,  
must not interact via the exchange of $W$ or $Z$ bosons, and is, therefore, sterile.  Second, we must address the fact that sterile neutrinos would be produced in the early universe through oscillations, thus affecting cosmological fits. Cosmological results are still not precise enough to make strong statements, although recent cosmological fits allow for a single non-interacting sterile neutrino \cite{Riess:2016jrr}.  However, it is also possible that, while sterile neutrinos may not interact weakly, they could have interactions through some new weaker-than-weak interactions that are yet to be observed.  In such models, the sterile neutrinos may not thermalize in the early universe, and so cosmological measurements will be insensitive to a sterile neutrino signal \cite{whitepaper}.   Third, the addition of the fourth mass state will affect $\beta$ decay.  A few-eV sterile neutrino will produce a kink in the $\beta$ decay electron spectrum, since the electron flavor state will have a small admixture associated with the high mass sterile state.  The upcoming KATRIN experiment could see a kink in the spectrum for masses greater than a few eV and mixings at the 0.1 level \cite{Formaggio:2011jg}.  Higher mass sterile neutrinos in the keV mass range can also affect the $\beta$ decay spectrum away from the endpoint and sensitivity to small mixings in the keV spectral range may also be possible \cite{Mertens:2014nha}.  Recent astrophysical measurements have shown hints of keV mass objects leading to renewed interest in these higher mass sterile neutrinos \cite{keVwhite}.

In this discussion, we will focus on neutrino oscillations.  This is the most immediately promising approach to observing evidence for sterile neutrinos, which show up through oscillations with frequency associated with the fourth mass state.   Following the arguments of Sec.~\ref{basics}, 
the optimal $L/E$ to study the additional fourth mass state indicated by LSND is of order 1 m/MeV, while for atmospheric oscillations it is 
 1000 km/GeV and for solar 10,000 km/GeV.  Since most neutrino beams are primarily $E<5$ GeV, this implies an $L<5$ km is required for studies of this 4th mass state, which is a much shorter baseline than is called for to study atmospheric and solar oscillations.   If one designs the experiment with ${\cal O}(1~{\rm m/MeV})$, then the sensitivities to $\Delta m^2_{solar}$ and $\Delta m^2_{atmospheric}$ are small.   One can make the approximation that  $\Delta m^2_{solar}=\Delta m^2_{atmospheric}=0$.   This is called the ``short baseline approximation.''  Invoking this approximation, we usually call the LSND-related mass splitting $\Delta m_{41}^2$.

Disappearance of an active flavor to a sterile flavor at large $\Delta m^2$ is direct evidence for these neutrinos.   Active-to-active appearance oscillations with a frequency associated with the third $\Delta m_{41}^2$ splitting is also possible.   
There are nine possible oscillations that can be observed with this oscillation frequency: $P_{\nu_e \rightarrow \nu_e}$,    $P_{\nu_\mu \rightarrow \nu_\mu}$,  $P_{\nu_\tau \rightarrow \nu_\tau}$,  $P_{\nu_e \rightarrow \nu_\mu}$ , $P_{\nu_\mu \rightarrow \nu_e}$ ,  $P_{\nu_e \rightarrow \nu_\tau}$, $P_{\nu_\tau \rightarrow \nu_e}$, $P_{\nu_\mu \rightarrow \nu_\tau}$ and $P_{\nu_\tau \rightarrow \nu_\mu}$.  All of these processes must occur with the same oscillation frequency for the model to be consistent.

The probabilities for disappearance and appearance oscillations are given by:   
\begin{equation}
P(\nu_{\alpha}\rightarrow\nu_{\beta})\simeq  4|U_{\alpha4}|^2 |U_{\beta4}|^2\sin^2(1.27\Delta m^2_{41}L/E)~, \label{1app}
\end{equation}
and
\begin{equation}
P(\nu_{\alpha}\rightarrow\nu_{\alpha})\simeq
1-4(1-|U_{\alpha4}|^2)|U_{\alpha4}|^2\sin^2(1.27\Delta m^2_{41}L/E)~. \label{disappeq2}
\end{equation}
Here we have used $\simeq$ for the relationships to explicitly note that we are employing the short baseline approximation and so have dropped  the small terms depending on the atmospheric and solar $\Delta m^2$ parameters.   Throughout the remaining text, we will just call this an equality.

From this, one can see that there are triplets of experiment-types, happening with the same oscillation frequency, that depend on pairs of the matrix elements--for example the set:
\begin{eqnarray}
P_{\nu_e \rightarrow \nu_e}&= & 1-4(1-|U_{e4}|^2)|U_{e4}|^2\sin^2(1.27\Delta m^2_{41}L/E)~, \label{PUe4}\\
P_{\nu_\mu \rightarrow \nu_\mu}&= & 1-4(1-|U_{\mu 4}|^2)|U_{\mu 4}|^2\sin^2(1.27\Delta m^2_{41}L/E)~, \label{PUmu4}\\
P_{\nu_{\mu}\rightarrow\nu_e}&= &  4|U_{e4}|^2 |U_{\mu4}|^2\sin^2(1.27\Delta m^2_{41}L/E).\label{PUe4mu4}
\end{eqnarray}
\noindent A consistent theory model requires signals in all three of these oscillation modes.   As we will see in Sec.~\ref{currentresults}, this is a serious problem for 3+1 models, at present.     

The above equations are often written to mimic the form of Eq.~\ref{prob}, for example
\begin{eqnarray}
P_{\nu_e \rightarrow \nu_e}&=& 1-\sin^2 2\theta_{ee}\sin^2(1.27\Delta m^2_{41}L/E), 
\label{Pthetaee} \\
P_{\nu_\mu \rightarrow \nu_\mu}&=& 1-\sin^2 2\theta_{\mu\mu}\sin^2(1.27\Delta m^2_{42}L/E), \label{Pthetamumu} \\
P_{\nu_{\mu}\rightarrow\nu_e}&=& \sin^2 2\theta_{e\mu}\sin^2(1.27\Delta m^2_{41}L/E) \label{Pthetaemu},
\end{eqnarray}
where, comparing to Eqs.~\ref{PUe4} to \ref{PUe4mu4}, results in the definitions:
\begin{eqnarray}
\sin^22\theta_{ee} &=& 4(1-|U_{e4}|^2)|U_{e4}|^2, \\
\sin^22\theta_{\mu \mu } &=& 4(1-|U_{\mu 4}|^2)|U_{\mu 4}|^2,\\
\sin^2 2\theta_{e\mu} &=& 4|U_{e4}|^2 |U_{\mu4}|^2.
\end{eqnarray}
A similar triplet of equations exists for the other flavor pairings.

The $U$-matrix can be thought of as a rotation matrix.    This leads to yet another notation that one finds in the literature.  The following definitions of oscillation probabilities can arise from using the rotation angles:
\begin{eqnarray}
P_{\nu_e \rightarrow \nu_e} &\simeq & 1-\sin^2\left(2 \theta _{14}\right) \, \sin ^2(1.27 \Delta m^2_{41} L/E), \label{connectee}\\
P_{\nu_\mu \rightarrow \nu_\mu} &\simeq & 1- \sin^2\left(2 \theta_{24}\right) \, \sin ^2(1.27 \Delta m^2_{41} L/E) \label{equ:pmm2} , \\
P_{\nu_\mu \rightarrow \nu_e}  &\simeq  & \frac{1}{4} \sin^2 \left( 2 \theta_{14} \right) \, \sin^2 \left( 2 \theta_{24} \right) \, \sin ^2(1.27 \Delta m^2_{41} L/E) \label{connectemu}.
\end{eqnarray}
and similar definitions follow for the other flavor pairs.

The above presents many definitions of the same probabilities.  We discuss these because all of these definitions are used in the literature.  One must take care not to confuse axes presenting mixing in one parameter space with another when comparing published results.   Another common source of confusion is between plots presented in $\sin^2 2\theta$ and $\sin^2 \theta$.  The reader must always look carefully at the choice of axis representing the mixing. 

Invoking $CPT$ as a good symmetry places an important requirement on any given oscillation triplet.
$CPT$ states that $P(\nu_\alpha \rightarrow \nu_\beta) = P(\bar \nu_\beta \rightarrow \bar \nu_\alpha)$. 
Consider the case where $\alpha$ and $\beta$ are both electron flavor.  
Then $P(\nu_e \rightarrow \nu_e) = P(\bar \nu_e \rightarrow \bar \nu_e)$.   Similarly for all flavors:  the oscillation probability for neutrinos and antineutrinos must be the same.    
In this case, 
\begin{enumerate}
\item  $CPT$ invariance requires that $\bar \nu_\mu \rightarrow \bar \nu_e$ oscillations
  and $\nu_e \rightarrow \nu_\mu$ oscillations must be identical.
\item    The probability for $\nu_e$ disappearance must be as large or larger than the probability
 for $\nu_e \rightarrow \nu_\mu$ oscillations, since this represents only one oscillation channel.
\item $CPT$ invariance requires that the probability of $\bar \nu_e$ disappearance be the
same as the probability for $\nu_e$ disappearance, as argued above.  
\end{enumerate}
This then leads to constraints on the relation between appearance and disappearance measurements.  For example, if 
the sensitive region of a $\bar\nu_e$ disappearance experiment
entirely covers the allowed parameter space associated with another experiments $\bar \nu_\mu \rightarrow
\bar \nu_e$ appearance signal region,  then either the disappearance experiment must see a signal or the two experiments are incompatible.  This method of testing appearance signals will represent an interesting line of attack for studying sterile neutrino models, as discussed later.

Lastly, it is important to point out that if one invokes the short baseline approximation, then there will be no $CP$ violation in the 3+1 model.   $CP$ violation would introduce a difference in the neutrino and antineutrino appearance probabilities in a given channel.   
$CP$ violating terms come about as an interference between two or more oscillation frequencies, hence two or more $\Delta m^2$ values that are relatively close in magnitude are required for an effect to be observed.  In a 3+1 model that makes use of the short baseline approximation, there is only one non-zero $\Delta m^2$ and, thus, no $CP$ violation in the model.  

\subsection{3+2 and 3+3 \label{twoandthree}}

While the minimal extension to a three neutrino model is to add a single sterile neutrino,  in principle, one might expect three sterile neutrinos, reflecting the three-generation structure of the Standard Model.  This is called a 3+3 model.   In practice, however, experiments may only have sensitivity to two of the three new neutrinos, as we discuss in Sec.~\ref{globfits}.  In this case, we must fit to a 3+2 model.

A 3+2 model, which is a five mass-state model, assumes the short baseline approximation for the three lowest mass states, and two distinct mass splittings, $\Delta m^2_{41}$ and $\Delta m^2_{51}$.  Note that $\Delta m^2_{54}=\Delta m^2_{51}-\Delta m^2_{41}$. The corresponding 3+2 mixing matrix is an extension of Eq.~\ref{4mixmx} to a 5$\times$5 matrix.    To simply the appearance equation below, we will use notation where
\begin{equation}
|U_{\alpha i \beta j}| = |U_{\alpha i}|||U_{\beta j}|, \label{Uabbrev}
\end{equation}
where $i$ and $j$ refer to the mass states, 
and 
\begin{equation}
\Delta_{ij} = \Delta m^2_{ij}. \label{dm2abbrev}
\end{equation}

For a 3+2 oscillation model, the appearance probability is then given by 
\begin{eqnarray}
P^{\rm 3+2}_{\nu_{\alpha}\rightarrow\nu_{\beta}}&\simeq& -4|U_{\alpha5\beta5}||U_{\alpha4\beta4}|\cos\phi_{54}\sin^2(1.27\Delta_{54} L/E) \nonumber \\
&~& +4(|U_{\alpha4\beta4}|+|U_{\alpha5\beta5}|\cos\phi_{54})|U_{\alpha4\beta4}|\sin^2(1.27 \Delta_{41} L/E) \nonumber\\
&~& +4(|U_{\alpha4\beta4}|\cos\phi_{54}+|U_{\alpha5\beta5}|)|U_{\alpha5\beta5}|\sin^2(1.27\Delta_{51} L/E) \nonumber\\
&~&+2|U_{\beta5\alpha5}||U_{\beta4\alpha4}|\sin\phi_{54}\sin(2.53\Delta_{54} L/E) \nonumber\\
&~&+2(|U_{\alpha5\beta5}|\sin\phi_{54})|U_{\alpha4\beta4}|\sin(2.53\Delta_{41} L/E) \nonumber\\
&~&+2(-|U_{\alpha4\beta4}|\sin\phi_{54})|U_{\alpha5\beta5}|\sin(2.53\Delta_{51} L/E).\label{2app}
\end{eqnarray}
In this equation, $\phi$ is a 
$CP$ phase is given by
\begin{equation}
\label{cpvphase}
\phi_{54}=\mathrm{arg}(U_{e5}U_{\mu5}^*U_{e4}^*U_{\mu4}).
\end{equation}
The terms which depend on $\cos\phi_{54}$ are $CP$ conserving since they do not change sign in going from neutrino to antineutrino oscillations, while those that depend on $\sin\phi_{54}$ are $CP$ violating.  $CP$ effects only arise in appearance experiments. 

The 3+2 disappearance probability, which is simpler and so we will not use abbreviations, has no dependence on $\phi_{54}$:
\begin{eqnarray}
P^{\rm 3+2}_{\nu_{\alpha}\rightarrow\nu_{\alpha}}&\simeq&
1-4|U_{\alpha4}|^2|U_{\alpha5}|^2\sin^2(1.27\Delta m^2_{54}L/E)\nonumber \\
&& -4(1-|U_{\alpha4}|^2-|U_{\alpha5}|^2)(|U_{\alpha4}|^2\sin^2(1.27\Delta m^2_{41}L/E) \nonumber \\
&& +|U_{\alpha5}|^2\sin^2(1.27\Delta m^2_{51}L/E))~.  \label{disappeq2}
\end{eqnarray}
In other words,  disappearance experiments have no $CP$ violating effects.   Only appearance experiments can potentially demonstrate $CP$ violation.

In the $\sim1$ eV$^2$ region, short baseline experiments do not constrain $|U_{\tau 4}|$ because the neutrino energy required for the optimal $L/E$ is well below the $\tau$ production threshold of 3.48 GeV for the incident neutrino energy needed to produce a $\tau$-lepton.     Therefore, the global fits to the short baseline data sets are constraining 
$\Delta m^2_{41}$, $|U_{e4}|$, $|U_{\mu 4}|$, $\Delta m^2_{51}$, $|U_{e5}|$, $|U_{\mu 5}|$, and the CP parameter $\phi_{54}$.

The 3+2 model has introduced more flexibility into the global fits in two ways.  The first is to introduce more parameters.  The second is through allowing for $CP$ violation, which makes the appearance oscillation probabilities for neutrinos and antineutrinos have different dependencies on the mixing matrix elements. In other words,  $P_{\nu_\mu \rightarrow \nu_e}$ can be different from $P_{\bar\nu_\mu \rightarrow \bar\nu_e}$ even with the same mixing matrix.

One can take the next step, expanding to a 3+3 model, with a 6$\times$6 mixing matrix.  The appearance probability, expressed using the notation in Eqs.~\ref{Uabbrev} and \ref{dm2abbrev}, is then given by:
\begin{eqnarray}
P^{\rm 3+3}_{\nu_{\alpha}\rightarrow\nu_{\beta}}&\simeq& 
-4|U_{\alpha5\beta5}||U_{\alpha4\beta4}|\cos\phi_{54}\sin^2(1.27\Delta_{54}L/E) \nonumber\\
&~& -4|U_{\alpha6\beta6}||U_{\alpha4\beta4}|\cos\phi_{64}\sin^2(1.27\Delta_{64}L/E) \nonumber\\
&~& -4|U_{\alpha5\beta5}||U_{\alpha6\beta6}|\cos\phi_{65}\sin^2(1.27\Delta_{65}L/E) \nonumber\\
&~& +4(|U_{\alpha4\beta4}|+|U_{\alpha5\beta5}|\cos\phi_{54}+|U_{\alpha6\beta6}|\cos\phi_{64})|U_{\alpha4\beta4}|\sin^2(1.27\Delta_{41}L/E) \nonumber\\
&~& +4(|U_{\alpha4\beta4}|\cos\phi_{54}+|U_{\alpha5\beta5}|+|U_{\alpha6\beta6}|\cos\phi_{65})|U_{\alpha5\beta5}|\sin^2(1.27\Delta_{51}L/E) \nonumber\\
&~& +4(|U_{\alpha4\beta4}|\cos\phi_{64}+|U_{\alpha5\beta5}|\cos\phi_{65}+|U_{\alpha6\beta6}|)|U_{\alpha6\beta6}|\sin^2(1.27\Delta_{61}L/E) \nonumber\\
&~&+2|U_{\beta5\alpha5}||U_{\beta4\alpha4}|\sin\phi_{54}\sin(2.53\Delta_{54}L/E) \nonumber\\
&~& +2|U_{\beta6\alpha6}||U_{\beta4\alpha4}|\sin\phi_{64}\sin(2.53\Delta_{64}L/E) \nonumber\\
&~&+2|U_{\beta6\alpha6}||U_{\beta5\alpha5}|\sin\phi_{65}\sin(2.53\Delta_{65}L/E) \nonumber\\
&~&+2(|U_{\alpha5\beta5}|\sin\phi_{54}+|U_{\alpha6\beta6}|\sin\phi_{64})|U_{\alpha4\beta4}|\sin(2.53\Delta_{41}L/E) \nonumber\\
&~&+2(-|U_{\alpha4\beta4}|\sin\phi_{54}+|U_{\alpha6\beta6}|\sin\phi_{65})|U_{\alpha5\beta5}|\sin(2.53\Delta_{51}L/E) \nonumber\\
&~&+2(-|U_{\alpha4\beta4}|\sin\phi_{64}-|U_{\alpha5\beta5}|\sin\phi_{65})|U_{\alpha6\beta6}|\sin(2.53\Delta_{61}L/E).\label{3app}
\end{eqnarray}
The disappearance equation is much simpler, and so, without abbreviations is:
\begin{eqnarray}
P^{\rm 3+3}_{\nu_{\alpha}\rightarrow\nu_{\alpha}}&\simeq&
1-4|U_{\alpha4}|^2|U_{\alpha5}|^2\sin^2(1.27\Delta m^2_{54}L/E)\nonumber \\
&~&-4|U_{\alpha4}|^2|U_{\alpha6}|^2\sin^2(1.27\Delta m^2_{64}L/E)-4|U_{\alpha5}|^2|U_{\alpha6}|^2\sin^2(1.27\Delta m^2_{65}L/E) \nonumber \\
&~&-4(1-|U_{\alpha4}|^2-|U_{\alpha5}|^2-|U_{\alpha6}|^2)(|U_{\alpha4}|^2\sin^2(1.27\Delta m^2_{41}L/E) \nonumber \\
&~&+|U_{\alpha5}|^2\sin^2(1.27\Delta
m^2_{51})+|U_{\alpha6}|^2\sin^2(1.27\Delta m^2_{61}L/E)).  \label{disappeq}
\end{eqnarray}
The important point here is that, again,  $CP$ phases are only in the appearance formula.
With the introduction of another mass splitting, three $CP$ violating phases arise: 
\begin{eqnarray}
\label{cpvphase}
\phi_{54}=\mathrm{arg}(U_{e5}U_{\mu5}^*U_{e4}^*U_{\mu4})~, \\
\label{cpvphase2}
\phi_{64}=\mathrm{arg}(U_{e6}U_{\mu6}^*U_{e4}^*U_{\mu4})~,
\end{eqnarray}
and
\begin{eqnarray}
\label{cpvphase2}
\phi_{65}=\mathrm{arg}(U_{e6}U_{\mu6}^*U_{e5}^*U_{\mu5})~.
\end{eqnarray}

From these equations, one sees that the numbers of parameters in the appearance fits are three for 3+1 ($|U_{\alpha4}|, |U_{\beta4}|$ and $\Delta m^2_{41}$), while the disappearance fits each have one less mixing matrix element to fit.    For 3+2 the number of parameters goes up to seven for appearance, having added in $|U_{\alpha 5}|$, $|U_{\beta5}|$, $\Delta m^2_{51}$ and the $CP$ violating parameter $\phi_{54}$.  Again, for disappearance fits, there is only one mixing matrix connecting to the 5th mass state, and no $CP$ violation.    Lastly,  for 3+3 appearance, the number of parameters rises to 12 for appearance, now including 
$|U_{\alpha 6}|$, $|U_{\beta6}|$, $\Delta m^2_{61}$, and two additional $CP$ violating parameters.   Thus, as expected, adding sterile neutrino states does add parameters, potentially improving fits, but the number of additional parameters turns out to be relatively small compared to the number of data sets, and far less than the number of bins in the full fit. 

The 3+3 model is arguably the most natural.   However, it has a very large number of parameters, thus requiring more data sets to perform a useful  fit.  It should be noted that, in the case where $\Delta m^2_{61}$ is very large, the typical $L/E$ values for short baseline experiments are comparatively low so that the terms depending on this splitting will have rapid oscillations, which leads to a flat contribution.   If the mixing angles are also small, then this offset will not have a strong affect on the fits and the system will reduce to an effective 3+2 model.

\subsection{Comparing experimental results in 3+1 models \label{LE}}

Experimental searches for sterile neutrinos in either appearance or disappearance mode rely on comparing the measured event rate in some channel to a prediction including backgrounds and a possible sterile neutrino signal.  It is difficult to compare these event rates directly since each experiment has differences in setups, efficiencies, resolutions, and backgrounds.  Typically, experiments rely on simulations to do this comparison and then do fits to extract allowed regions or limits in terms of oscillation parameters.  
For a 3+1 model, these regions can be displayed for a given experiment as allowed or excluded regions in the $\Delta m^2$ {\it vs.} $\sin^22\theta$ plane.  The LSND results, discussed in Section \ref{2examples}, are an example of this procedure.

Experimental results presented in this $\Delta m^2$ {\it vs.} $\sin^22\theta$ parameter space can be easily compared via the overlap of regions from the various experiments. This method has the advantage of incorporating all the information from the given experiments and putting the results on a common footing that can be rigorously compared.  In particular, the distribution of ``true'' neutrino energies for any given``reconstructed'' neutrino energy can be used to estimate the oscillation regions, and the systematic uncertainties and correlations associated with neutrino flux, backgrounds, and reconstruction at different energies can be correctly applied. Comparing experiments through this type of oscillation phenomenology is the only rigorous method to compare and combine experiments and is the basis of the global fits to multi-experiment results discussed in Section \ref{globfits}.  

\begin{figure}[p]
\vspace{-0.25in}
\centerline{\includegraphics[width=0.6\textwidth]{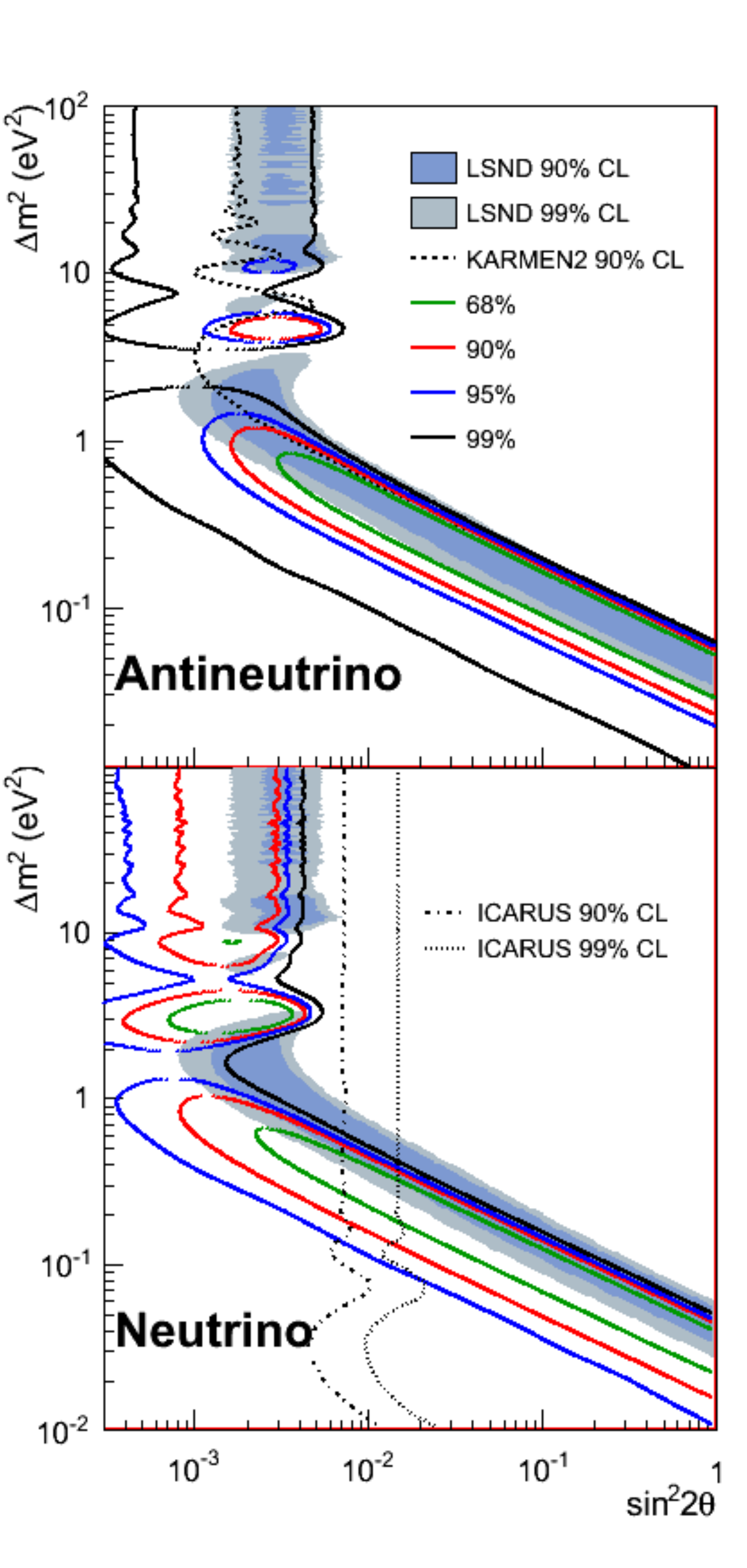}}
 \vspace{-0.3in}
\caption{MiniBooNE \cite{MBPRL} allowed regions in antineutrino mode (top) 
and neutrino mode (bottom) for events with
$E^{QE}_{\nu} > 200$ MeV within a two-neutrino oscillation model. 
Also shown are the ICARUS \cite{ICARUS} and KARMEN \cite{karmen} 
appearance limits for neutrinos and antineutrinos, respectively.
The shaded areas show the 90\% and 99\% C.L. LSND 
$\bar{\nu}_{\mu}\rightarrow\bar{\nu}_e$ allowed 
regions.}
\label{limitab}
\vspace{-0.1in}
\end{figure}

To provide an example, consider a comparison the results of the MiniBooNE experiment to LSND.
MiniBooNE is a muon-to-electron flavor appearance experiment that we will describe in section \ref{currentresults}, below.   It was designed to follow up on the LSND anomaly, running in both neutrino and antineutrino mode.  The $L/E$ of MiniBooNE was selected to allow coverage of the LSND allowed region.   MiniBooNE observed an anomalous excess consistent with oscillations in both running modes.  Results from the MiniBooNE experiment \cite{MBPRL} are shown in Fig.~\ref{limitab}.   The proper way to compare MiniBooNE to LSND is to overlay the two-neutrino oscillation allowed regions, as shown in this figure.

An alternative method to compare experiments \cite{arxivLE} that has been used in the literature is to calculate the observed oscillation probability in $L/E$ bins to exploit the expected dependence of Eq. \ref{prob}.  This is a poor choice, because, unfortunately, the true $L$ and true $E$ of an event cannot be determined due to experimental energy resolution and the smearing associated with a finite length neutrino source and event position detection.  So, a given $L/E$ bin has contributions from a range true $L/E$ values.  The measured oscillation probability also has uncertainties associated with the modeling of the backgrounds and predicted signal rate. These uncertainties will typically introduce correlations between the various measured data points.  
 
\begin{figure}[t]
\vspace{+0.1in}
\centerline{\includegraphics[angle=0, width=0.75\textwidth]{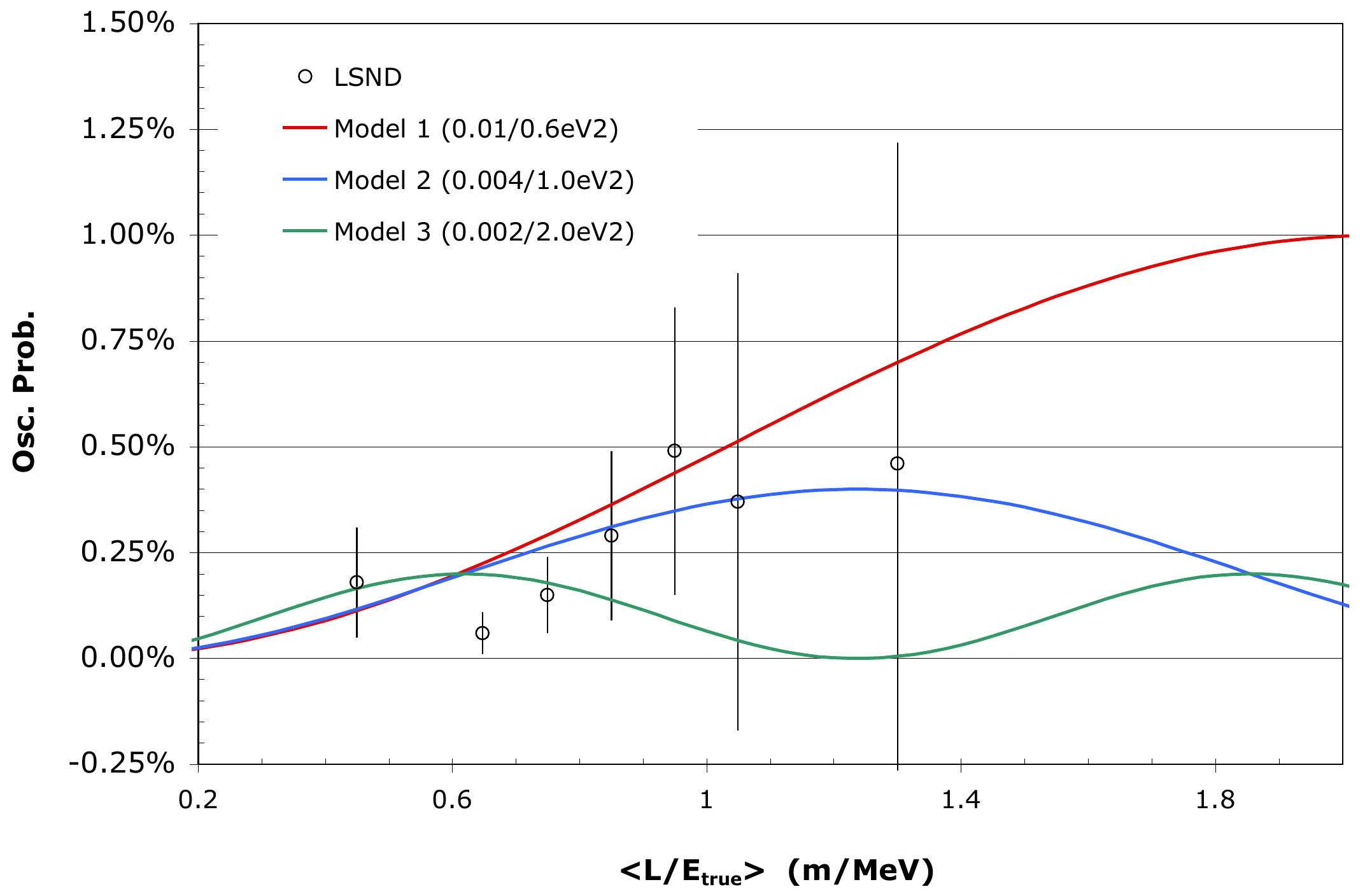}}
\vspace{-0.1in}
\caption{The LSND measured oscillation probability as a function of  the reconstructed $L/E$.   Three theoretical curves without any energy or flight path smearing are also shown for models with $\sin^22\theta/\Delta m^2 (eV^2) = 0.01/0.6, 0.004/1.0, $ and $0.002/2.0$. (From Ref.~\cite{Aguilar-Arevalo:2014xrr}.) } 
\label{LSND_LovrE}
\vspace{-0.1in}
\end{figure}

\begin{figure}[t]
\vspace{+0.1in}
\centerline{\includegraphics[angle=0, width=0.75\textwidth]{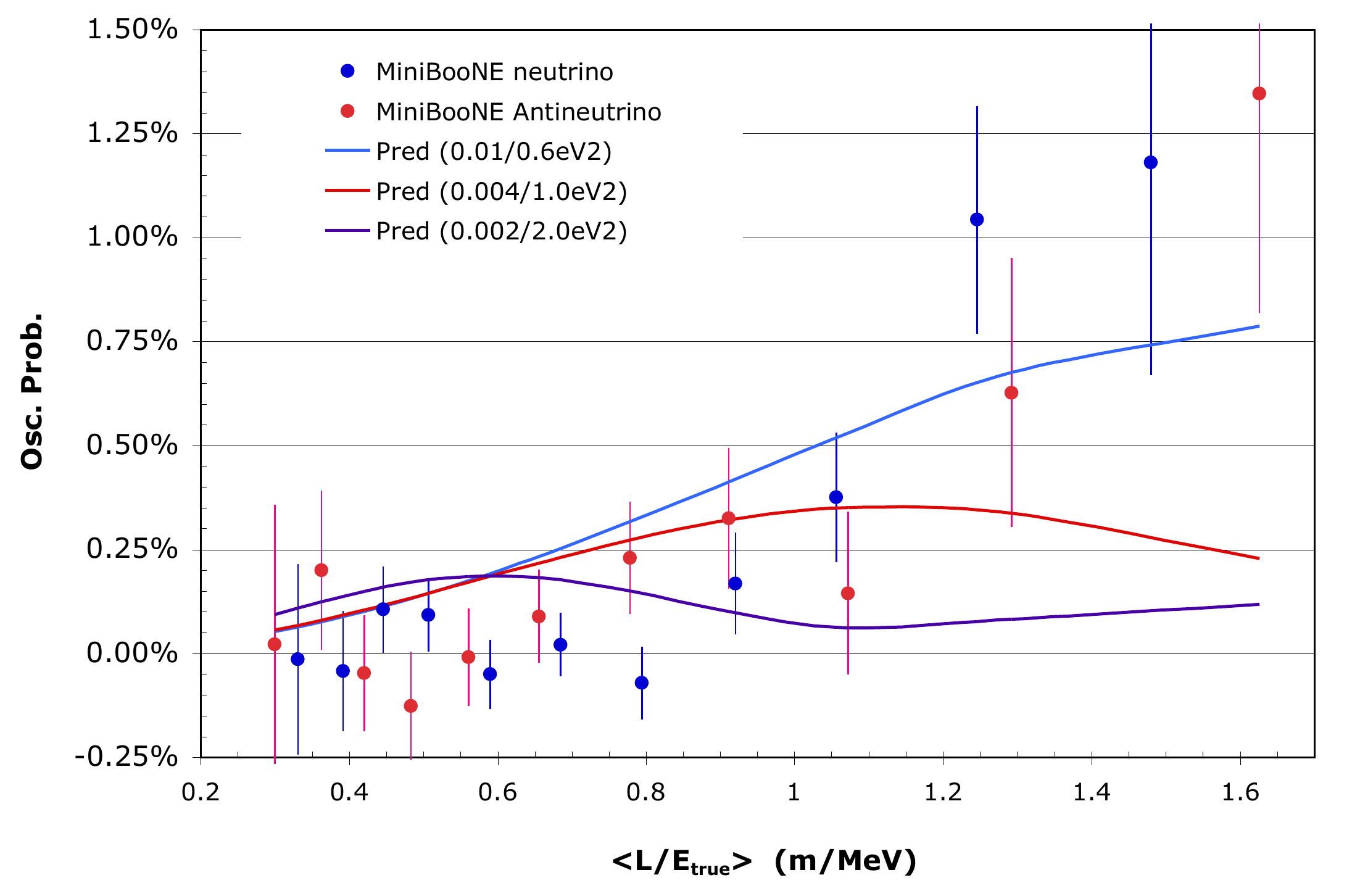}}
\vspace{-0.1in}
\caption{The  MiniBooNE $P_{osc}$ measurements as a function of $L/E$ for neutrino and antineutrino mode running.  
The curves are the predicted $P_{osc}$ versus $L/E$ with energy and flight path smearing.(From Ref.~\cite{Aguilar-Arevalo:2014xrr})} 
\label{MB_with_Pred}
\vspace{-0.1in}
\end{figure}

All of the above effects are likely to be experiment-dependent and, thus, render the $L/E$ method not a very robust comparison technique.  In order to apply the $L/E$ method to an actual experiment, one must, therefore, use a simulation as a tool to make corrections and assess measured values.  Simulated data is binned in measured $L/E$ and the predicted oscillation probability for a given set of oscillation model parameters is calculated. This procedure will be experiment-dependent and so an $L/E$ plot can only be made for a single experiment and used to compare the measured data to various oscillation models. Two example $L/E$ plots (from Ref.~\cite{Aguilar-Arevalo:2014xrr}) are given in Fig.~\ref{LSND_LovrE} for LSND and in Fig.~\ref{MB_with_Pred} for MiniBooNE $\nu$ and $\bar\nu$ results.  Because of the bin-size compared to the accuracy of the experiment, the LSND experimental measurement of $L/E$  does not need smearing corrections. However, it should be noted that the data points do have correlated uncertainties associated with the $P_{osc}$ measurement. For MiniBooNE, the smearing effects are significant and the procedure outlined above needs to be used.    
 
Lastly, care must be taken when comparing experiments that are measuring different, but related parameters.     Consider, for example, the triplet of measurements shown in Eqs. \ref{PUe4} to \ref{PUe4mu4}, which are the set of $\nu_e \rightarrow \nu_e$, $\nu_\mu \rightarrow \nu_\mu$ and $\nu_\mu \rightarrow \nu_e$.    In order to properly put all results from all three types of experiments onto the same plot, one has to globally fit all of the data to extract $|U_{e4}|$ and $|U_{\mu 4}|$ at every $\Delta m^2$, and use the correct values at each point to project into a given plane, for example the $\sin^2 2\theta_{e\mu}$ plane.    This is rarely done in the literature. Most of the time, the best fit values from some global fit are used to project allowed regions and limits, leading to some distortion of the curves.  It is important for readers to be aware of this and to consider such projected curves as qualitative rather than quantitative.

\subsection{Matter effects \label{matter}}

Up to this point, we have only considered vacuum oscillations.   This is appropriate for experiments where the neutrinos traverse short distances through matter of low density.  This is true of nearly all experiments that are sensitive to oscillations in the $\sim 1$ eV$^2$ region.   However there is an important recent exception:  the $\nu_\mu$ disappearance result from IceCube \cite{IcePRL}.

This analysis made use of a high-statistics sample of
ultra-high-energy atmospheric neutrino interactions in the 0.4 to 20 
TeV energy range.    The path-length is related to the angle of the incoming neutrinos, and those which are upward-going through the earth were used.
The 
energy and pathlength, given the size of the earth, results in $L/E \sim 1$
m/MeV, which is the vacuum-oscillation parameter range-of-interest for observed
short-baseline sterile
neutrino anomalies \cite{SBL2016}.   However the resolution of
the IceCube detector causes a vacuum-oscillation analysis to be
insensitive.     The strength of the IceCube result arises from a
matter-effect signature in IceCube \cite{mattereffectsice} that 
predicts  
a large deficit in the antineutrino 
flux for the up-going neutrinos that cross the Earth.   The
matter-magnified signal greatly
enhances the IceCube sensitivity to  a 3+1
models consistent with the observed short baseline anomalies.

To understand the source of the matter effects in sterile neutrino searches, let's first consider a known matter effect in the three neutrino model.   If active neutrinos traverse an environment with a high density of electrons, then this environment introduces an additional potential to the Hamiltonian that modifies the  $\nu_e$ flavor component.  The potential V$_{CC}$ is proportional to the Fermi coupling constant $G_f$ and the density of electrons $n_e$: 
\begin{equation}
V_{CC} = \sqrt{2} G_{f} n_e.
\end{equation}
This affects vacuum oscillations involving the $\nu_e$ content of the propagating neutrino.   Transitions of neutrino flavors within the sun, which has a very high density of electrons, are known to be affected by this potential \cite{mattereffectsSun}.

Analogously, in a four-neutrino model, as neutrinos pass through the denser regions of the earth, the propagation of the active flavor components that interact with matter will be affected while the sterile component will not.   This introduces an additional potential into the Hamiltonian, leading to a modification of the oscillation probability \cite{mattereffectsice}.   The result, for a 3+1 model, is a substantial predicted deficit for $\nu_\mu \rightarrow \nu_\mu$ in specific regions of energy and pathlength (angle) that is in the range of IceCube sensitivity.

\section{Existing sterile neutrino signals, hints, and limits \label{currentresults}}

There have been many searches for oscillations to sterile neutrinos.  Table~\ref{tab:explist} lists a number of the experiments that currently have results based on searches for vacuum oscillations ({\it i.e.} experiments without matter effects).   This table indicates the name of the experiment, search mode and whether the experiment has a closed contour at 95\% CL, which we will call a ``signal,'' or an open contour that includes the null, which we will call a ``limit.''   Note that this is an arbitrary choice, and that KARMEN/LSND XSEC \cite{ConradShaevitz}, MiniBooNE/SciBooNE $\nu$ \cite{Mahn:2011ea}, and CDHS \cite{CCFR84} have enclosed contours at 90\% CL but not at 95\% CL. None of the individual experimental results rise to the level of an ``observation,'' which is generally defined as a 5$\sigma$ signal.  

The earliest experiments were looking for oscillations among the three standard neutrinos in the $\Delta m^2$ region above 1 eV$^2$.  This was motivated by three-neutrino dark matter models that were viable at the time, but are now disfavored by cosmology \cite{oldDMmodels}.     Subsequently, the results of these experiments, such as CDHS and CCFR \cite{CCFR84}, were used to set limits on $\nu_\mu$ disappearance associated with possible oscillations to sterile neutrinos.  

\begin{table}[t]
\begin{center}
\begin{tabular}{|l|c|c|c|c|c|}
\hline
 & Process  & $\nu$/$\bar \nu$ & App/Dis &  Vac/Mat & Result \\ \hline
 Electron Neutrino Appearance & & & & & \\ \hline
LSND \cite{LSNDprimaryresult}  & $\bar \nu_\mu \rightarrow \bar \nu_e$ & $\bar \nu$ & App & Vac & Signal\\
MiniBooNE - $\nu$ \cite{miniboonelowe2, miniboonelowe1} & $\nu_\mu \rightarrow \nu_e$  & $\nu$ &
App & Vac & Signal\\
MiniBooNE - $\bar\nu$ \cite{MBPRL, nubarminiosc1}& $\bar \nu_\mu \rightarrow \bar
\nu_e$ & $\bar \nu$ & App & Vac & Signal\\
KARMEN \cite{karmen} & $\bar \nu_\mu \rightarrow \bar \nu_e$ & $\bar \nu$ & App & Vac & Limit\\
ICARUS \cite{ICARUS} & $\nu_\mu \rightarrow \nu_e$  & $\nu$ & App & Vac & Limit\\
NOMAD \cite{NOMAD1} & $\nu_\mu \rightarrow \nu_e$  & $\nu$ & App & Vac & Limit\\ \hline
Electron Neutrino Disappearance  & & & & & \\ \hline
Bugey and other reactors \cite{mention, BUGEY}& $\bar \nu_e \rightarrow \bar \nu_e$ & $\bar \nu$ & Dis & Vac & Signal \\
Gallium Exps. (SAGE\cite{SAGE3}, GALLEX \cite{GALLEX3}) & $\nu_e \rightarrow \nu_e$ & $\nu$ & Dis & Vac & Signal\\
KARMEN/LSND XSEC  \cite{ConradShaevitz} & $\nu_e \rightarrow \nu_e$ & $\nu$ & Dis & Vac & Limit\\ \hline
Muon Neutrino Disappearance   & & & & & \\ \hline
MiniBooNE/SciBooNE - $\nu$ \cite{Mahn:2011ea}& $\nu_\mu \rightarrow \nu_\mu$ & $\nu$
& Dis & Vac & Limit\\
MiniBooNE/SciBooNE - $\bar\nu$ \cite{MBSBantinu} & $\bar\nu_\mu \rightarrow \bar\nu_\mu$ & $\bar\nu$
& Dis & Vac & Limit\\
IceCube \cite{IceCube} & $\bar \nu_\mu \rightarrow \bar \nu_\mu$ & $\bar \nu$ & Dis & Mat & Limit\\
CCFR \cite{CCFR84} & $\nu_\mu \rightarrow \nu_\mu$  & $\nu$ & Dis & Vac & Limit\\
CDHS \cite{CDHS} & $\nu_\mu \rightarrow \nu_\mu$ & $\nu$ & Dis & Vac & Limit\\
MINOS \cite{MINOS2016,MINOSCC2}& $\nu_\mu \rightarrow \nu_\mu$ & $\bar \nu$ & Dis & Vac & Limit\\
\hline
\end{tabular}
\caption{Oscillation experiments with sensitivity to sterile neutrino oscillations in the $\Delta m^2$ region from 0.01 to 10 eV$^2$. The experiments are identified as appearance (App) or disappearance  (Dis).    They are also identified as ``Vac,'' for vacuum oscillations, as occur in short baseline experiments, and ``Mat,'' for matter dependence, which is at present unique to IceCube.   The results of each experiment are categorized either as an observation of a ``signal'' if there is a closed contour at 95\% CL,  or a ``limit'' otherwise.  See text for discussion.}
\label{tab:explist}
\end{center}
\end{table}

One of the first dedicated experiments searching for sterile neutrino oscillations was the LSND experiment at Los Alamos, which has been described previously in Section \ref{2examples}. Fig.~\ref{LSNDsignal} shows the 3.8$\sigma$ event excess of the data over the background, which fits well with the energy distribution expected for oscillations.  Fitting this data to a 3+1 oscillation models yields the allowed region shown in Fig.~\ref{fig:LSNDMBSB}.  The LSND $\nu_e$ appearance signal has prompted many follow-up experiments in all the possible channels including $\nu_e$ appearance, $\nu_e$ disappearance and $\nu_\mu$ disappearance in the $\Delta m^2$ region around 1 eV$^2$.

\begin{figure}[t]
  \centering
  \begin{minipage}[b]{0.48\textwidth}
  \includegraphics[width=\linewidth]{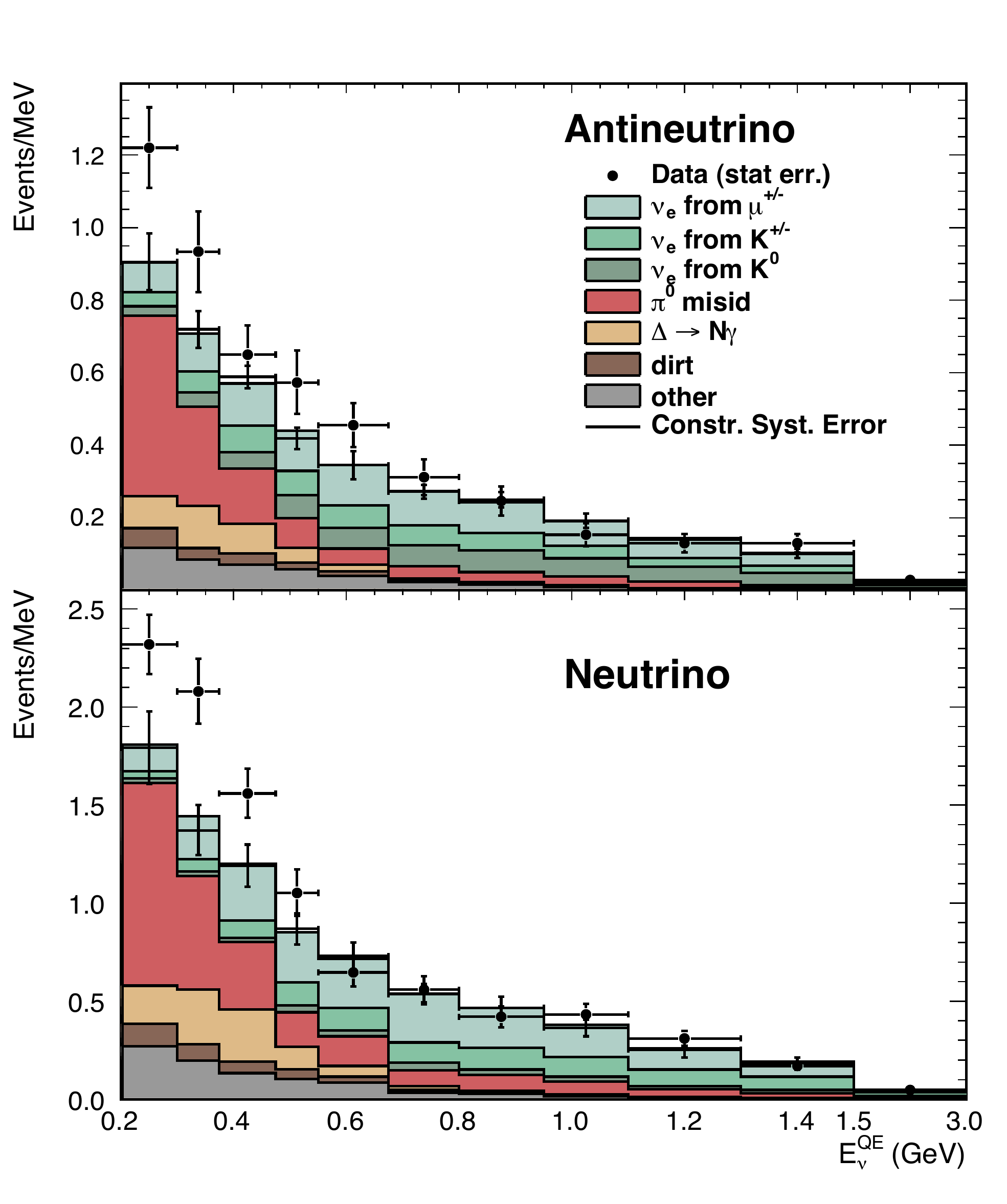}
    \caption{The antineutrino mode (top) and neutrino mode (bottom)  $E_\nu^{QE}$ distributions 
for ${\nu}_e$ CCQE data (points with statistical errors) and background (histogram with systematic errors). (From Ref.~\cite{MBPRL}.)}
  \label{MBevents}
  \end{minipage}
  \hfill
  \begin{minipage}[b]{0.48\textwidth}
    \includegraphics[width=\linewidth]{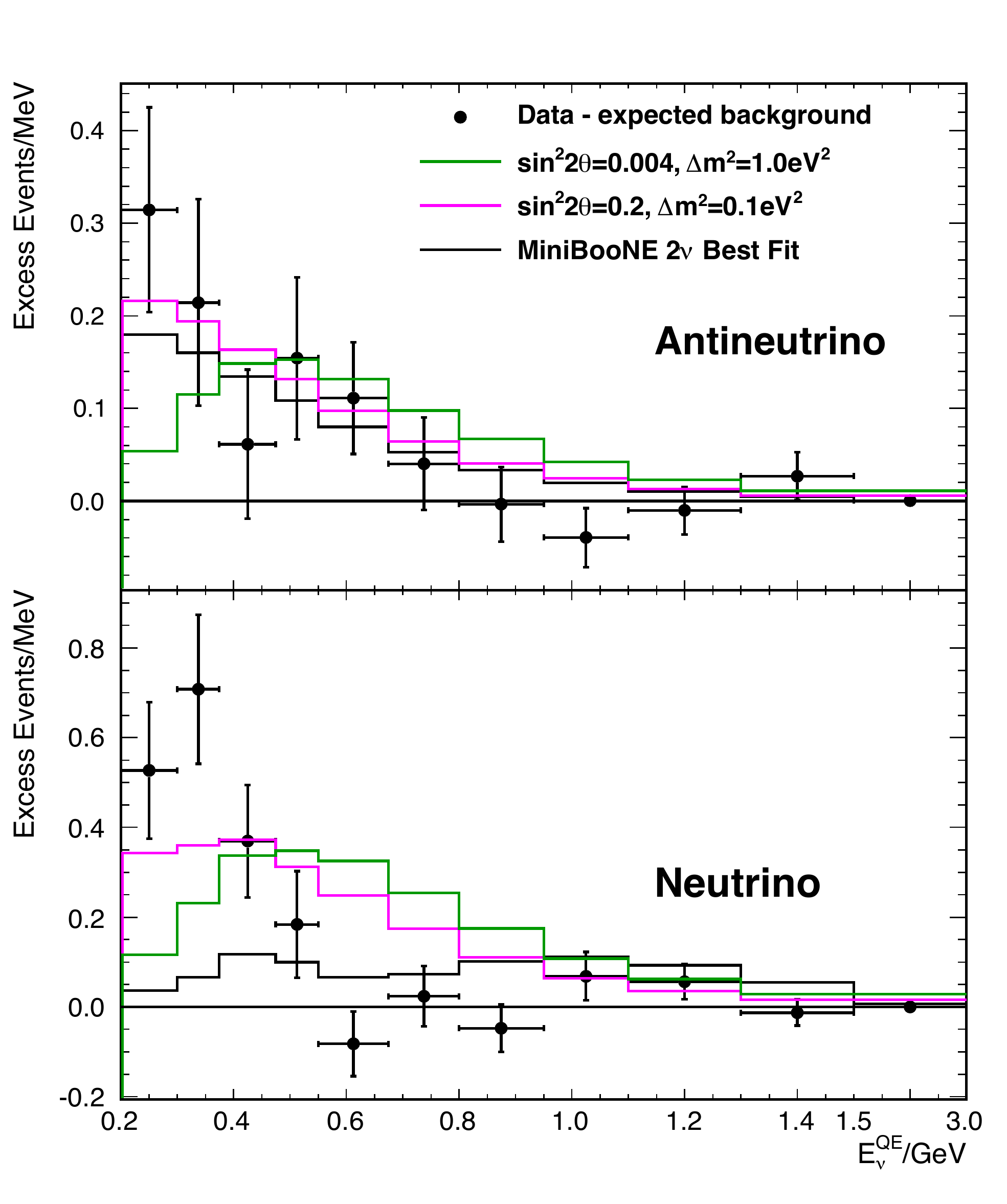}
    \caption{The antineutrino mode (top) and neutrino mode (bottom)
event excesses as a function of $E_\nu^{QE}$. (Error bars include
both the statistical and systematic uncertainties.) Also shown are the expectations from the best two-neutrino fit for each mode and 
for two example sets of oscillation parameters (From Ref.~\cite{MBPRL}.)}
  \label{MBexcess}
  \end{minipage}
    \label{MiniBooNE}
\end{figure}

As discussed previously, the MiniBooNE experiment was designed to test the LSND oscillation signal by searching for oscillations with a much different experimental setup but holding the $L/E$ value close to the LSND value.  For MiniBooNE, $L/E$ averaged around (540 m/ 600 MeV) whereas for LSND the value averaged (35 m/ 40 MeV).  MiniBooNE could run in both neutrino and antineutrino mode so could search for both $\nu_\mu \rightarrow \nu_e$ and $\bar\nu_\mu \rightarrow \bar\nu_e$ oscillations.  The systematic uncertainties and backgrounds were also much different than LSND.  MiniBooNE had higher relative backgrounds, but developed direct data techniques to constrain the impact on the oscillation analysis.  Fig.~\ref{MBevents} shows that the data compared to the expected background indicates a clear excess in both the  $\nu$ and $\bar\nu$ channels.  This excess is then displayed in Fig.~\ref{MBexcess} where the data minus the background is compared to several 3+1 oscillation models. Notice that when one background subtracts, one can end up with negative values for the oscillation excess--a point that often confuses even seasoned physicists. 

The antineutrino excess agrees well with several of the oscillation models but the neutrino data has an excess at low energy above any of the models.  This is often referred to as the MiniBooNE low-energy excess and is one of the problems with interpreting the measurements as indications of sterile neutrino oscillations with a simple 3+1 model.  The low-energy region is also where the $\gamma$-ray backgrounds are largest from, for example, ``$\pi^0$ misid'' since the detector did not have the capability to separate electron from $\gamma$ events.  MiniBooNE used direct measurements from the neutrino data to constrain these backgrounds to much below the excess level as displayed by the error bars in Fig.~\ref{MBevents}, bottom.  Nevertheless, the low-energy excess could be from some new type of neutrino process that produces single $\gamma$ rays and future experiments should have the capability to separate $\gamma$s from electrons.  

\begin{figure}[t]
  \centering
  \includegraphics[width=0.6\textwidth]{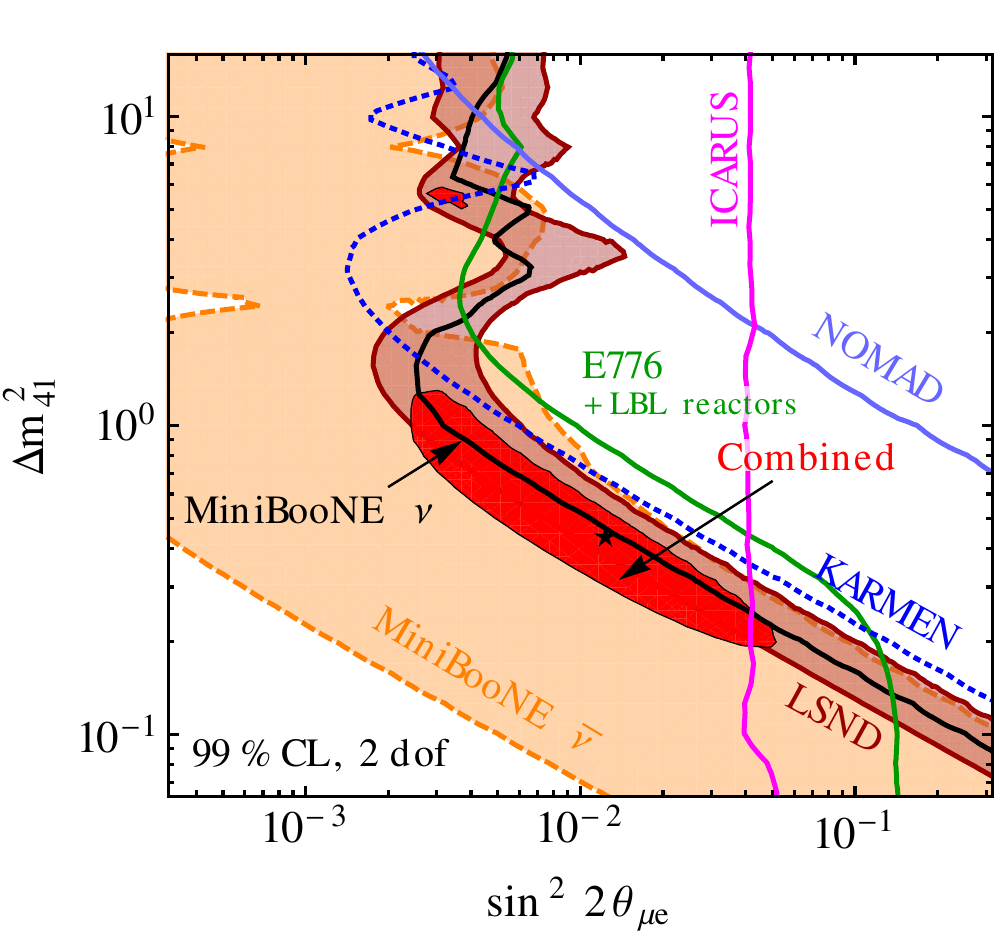}
  \caption{Allowed regions and upper bounds at 99\%~CL (2~dof) for
    muon to electron flavor appearance experiments
    in the 3+1 scheme. The red region corresponds to a combined fit of all
    $\nu_\mu \rightarrow \nu_e$ appearance data sets, with the star indicating the best fit point. \label{nue_app_exps}
    (From Ref.~\cite{Kopp_global}) 
    }
\end{figure}

The KARMEN \cite{karmen}, ICARUS \cite{ICARUS}, and NOMAD \cite{NOMAD1} experiments also searched for $\nu_\mu\rightarrow\nu_e$ appearance oscillations setting limits, but not with the sensitivity to fully exclude the MiniBooNE and LSND signals.  As a summary of the current situation, the allowed regions and limits for $\nu_\mu\rightarrow \nu_e$ appearance from 3+1 model fits to the various data sets are shown in Fig.~\ref{nue_app_exps}.  The red region in the figure shows the allowed region for a combined fit indicating that there are 3+1 oscillation parameter sets that are compatible with all the appearance data.

In a 3+1 model, $\nu_e$ appearance can only come about if both $|U_{e4}|$ and $|U_{\mu4}|$ are non-zero as shown in Eq.~\ref{PUe4mu4}.  Non-zero values for these elements implies that there must be both $\nu_e$ disappearance and $\nu_\mu$ disappearance, which leads also to consider the current results in those channels. 

Two of the $\nu_e$ disappearance searches have seen signals.  One is a set of results from beams produced by reactors producing antineutrinos. The other comes from studies of very intense radioactive sources producing neutrinos.

\begin{figure}[t]
  \includegraphics[width=1.1\textwidth]{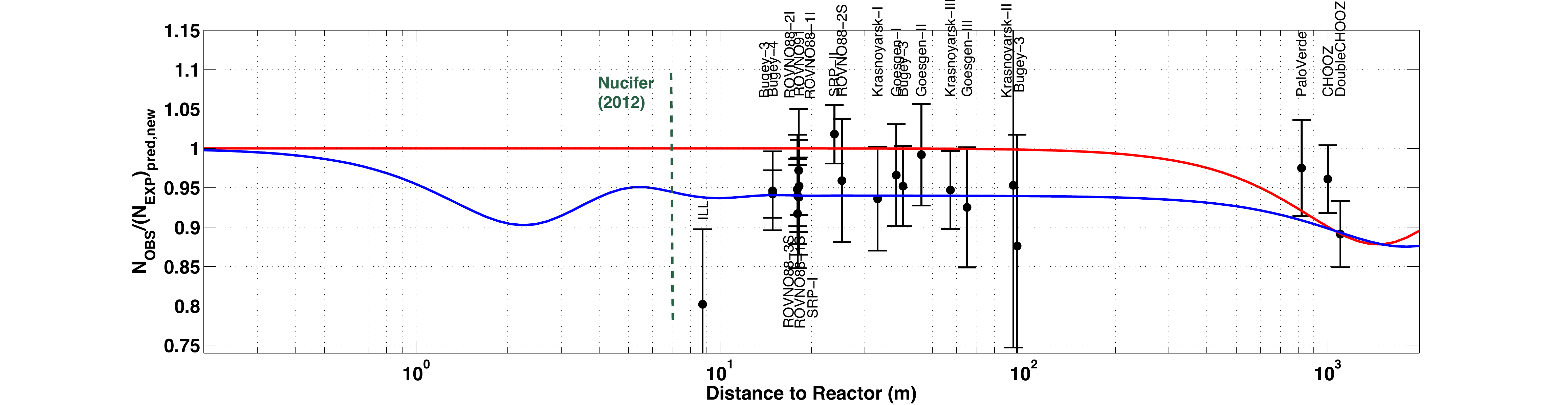}
  \caption{\label{raadistance} Short baseline reactor antineutrino anomaly.
    The experimental results are compared to the prediction without
    oscillations, taking into account the new antineutrino spectra, the
    corrections of the neutron mean lifetime, and the off-equilibrium effects.
    As an illustration, the red line shows a 3 active
    neutrino mixing solution and the blue line displays a solution including a
    new neutrino mass state, such as $|\Delta m_4^2 | \approx $ 1 eV$^2$ and
    $\sin^22\theta$=0.12. (From Ref.~\cite{whitepaper})}
\end{figure}

The first result, referred to as the ``Reactor Neutrino Anomaly''  \cite{mention, BUGEY}, uses the measured rate of $\bar\nu_e$ events detected from reactors as compared to prediction.  When one convolutes the reactor spectrum with the inverse beta decay ($\bar\nu_e+p \rightarrow e^+ + n)$ cross section, one obtains events with neutrino energies between 1.8 MeV and about 8 MeV, with a peak at about 3 MeV.
The absolute prediction of this flux is difficult to accomplish because of the complicated production mechanism of $\bar\nu_e$'s from the beta-decay of reactor fission fragments that must be modeled.  Recently, several experiments have observed an unmodeled bump in the higher energy part of the spectrum indicating some problem with the reactor model \cite{reactorbump}.  
For $|\Delta m^2|$ values near 1 eV$^2$, given the 3 MeV peak neutrino energy, the first oscillation maximum is near 3 m from the reactor.   Experiments near to the reactors are called ``short baseline reactor experiments.'' For $L$ values  $> \mathcal{O}(10)$ m, the probability should reach the fast-oscillation limiting value of $Prob=\frac{1}{2}\times \sin^22\theta$.     Experiments in this distance range from the reactor, and extending up to about a kilometer or more,  are called ``long baseline.''
Previous reactor measurements with L values in the 10 m to 100 m range have measured ratios to prediction less than one that average to $0.933\pm0.021$ as shown in Fig.~\ref{raadistance}.  This difference from $1.0$, the reactor anomaly, is a possible 3.2$\sigma$ indication of $\bar\nu_e$ disappearance.  

\begin{figure}[t]
  \centering
  \includegraphics[width=0.70\textwidth]{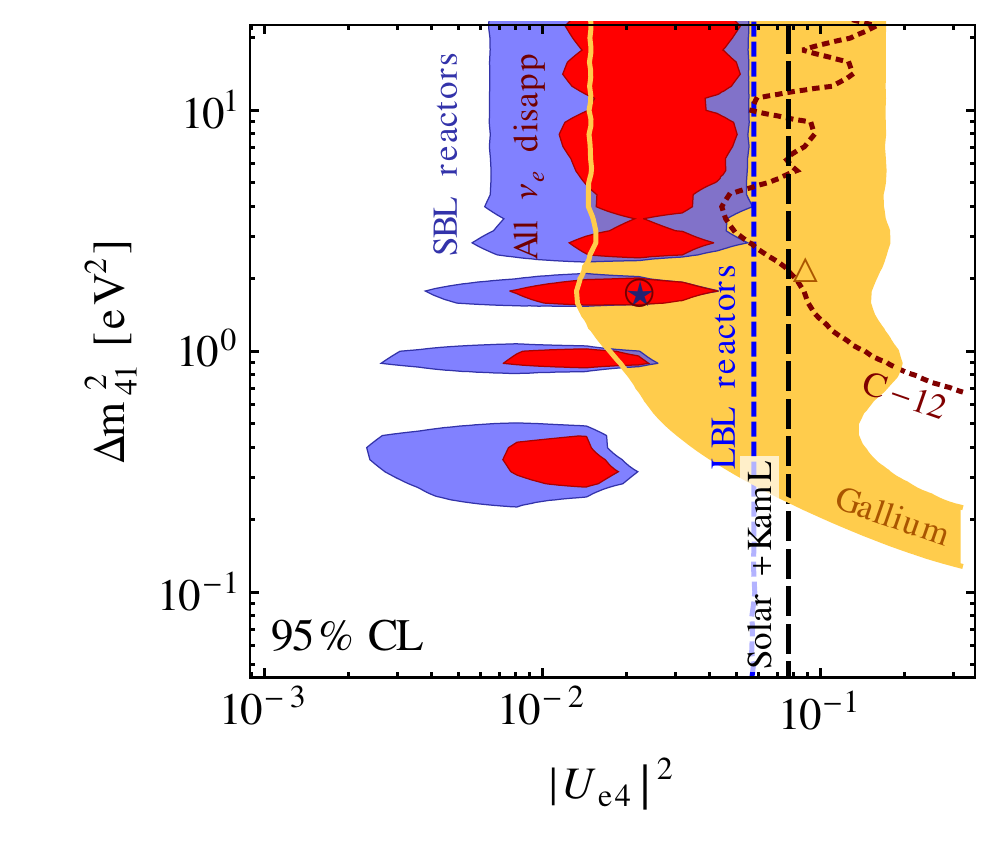}
  \caption{Allowed regions at 95\%~CL (2 dof) for 3+1
    oscillations. Regions shown are short-baseline reactor data (blue shaded), Gallium
    radioactive source data (orange shaded), $\nu_e$ disappearance
    constraints from comparing $\nu_e-$C-12 scattering data from LSND
    and KARMEN, called LSND-KARMEN XSEC in Table 1.1 (dark red dashed), long-baseline reactor data
(blue short-dashed),  and data from solar neutrino experiments, including the KamLAND result which is sensitive to solar oscillations  (black long-dashed). The red
    shaded region is the combined region from all of these $\nu_e$ and
    $\bar\nu_e$ disappearance limits.\label{nue_global} Note: $|U_{e4}|^2 \approx \sin^22\theta_{ee}/4$. (From Ref.~\cite{Kopp_global})}
\end{figure}

A second electron neutrino disappearance signal is associated with the measured rate of $\nu_e$'s produced by sources in the PBq ($10^{15}$ decays/second) range.  These are extremely hot sources--for comparison, the potassium decays that naturally occur in the human body are at the kBq level.     These sources were used in the ``Gallium experiments,'' GALLEX \cite{GALLEX3} and SAGE \cite{SAGE3}, which were solar neutrino detectors.  Again the measured rate is compared to a prediction and a deficit is found at the 2.9$\sigma$ level with $R_{obs}/R_{pred} = 0.84\pm0.05$.  This is referred to as the ``Gallium Anomaly'' since both detectors employ gallium in the detection medium.  The combination of the Reactor and Gallium anomalies along with other $\nu_e$ disappearance measurements leads to the allowed regions given in Fig.~\ref{nue_global}.  This collection of $\nu_e$ data are well fit by the 3+1 neutrino hypothesis, while the no-oscillation hypothesis is disfavored at 99.97\%~C.L (3.6 $\sigma$).

For $\nu_\mu$ disappearance, the present situation is quite different.  Currently there are no signals of sterile oscillations in this channel, using our 95\% CL definition.  
Recall that in a 3+1 model, the disappearance and appearance oscillation channels are not independent and coupled through the $|U_{e 4}|$ and
$|U_{\mu 4}|$ as shown in Eq.~\ref{PUe4mu4}.  These equations lead to the approximate relation between the effective mixing angles given by \begin{equation}\label{3+1relat}
  \sin^22\theta_{\mu e}
  \approx \frac{1}{4} \sin^22\theta_{ee}\sin^22\theta_{\mu\mu} \,.
\end{equation}
Thus the $\nu_e$ disappearance and $\nu_\mu \rightarrow \nu_e$ appearance signals described above lead to a prediction for $\nu_\mu$ disappearance. If no disappearance is seen in that range, with high confidence level, then the model is excluded.

\begin{figure}[t]
  \centering
\includegraphics[width=0.75\textwidth]{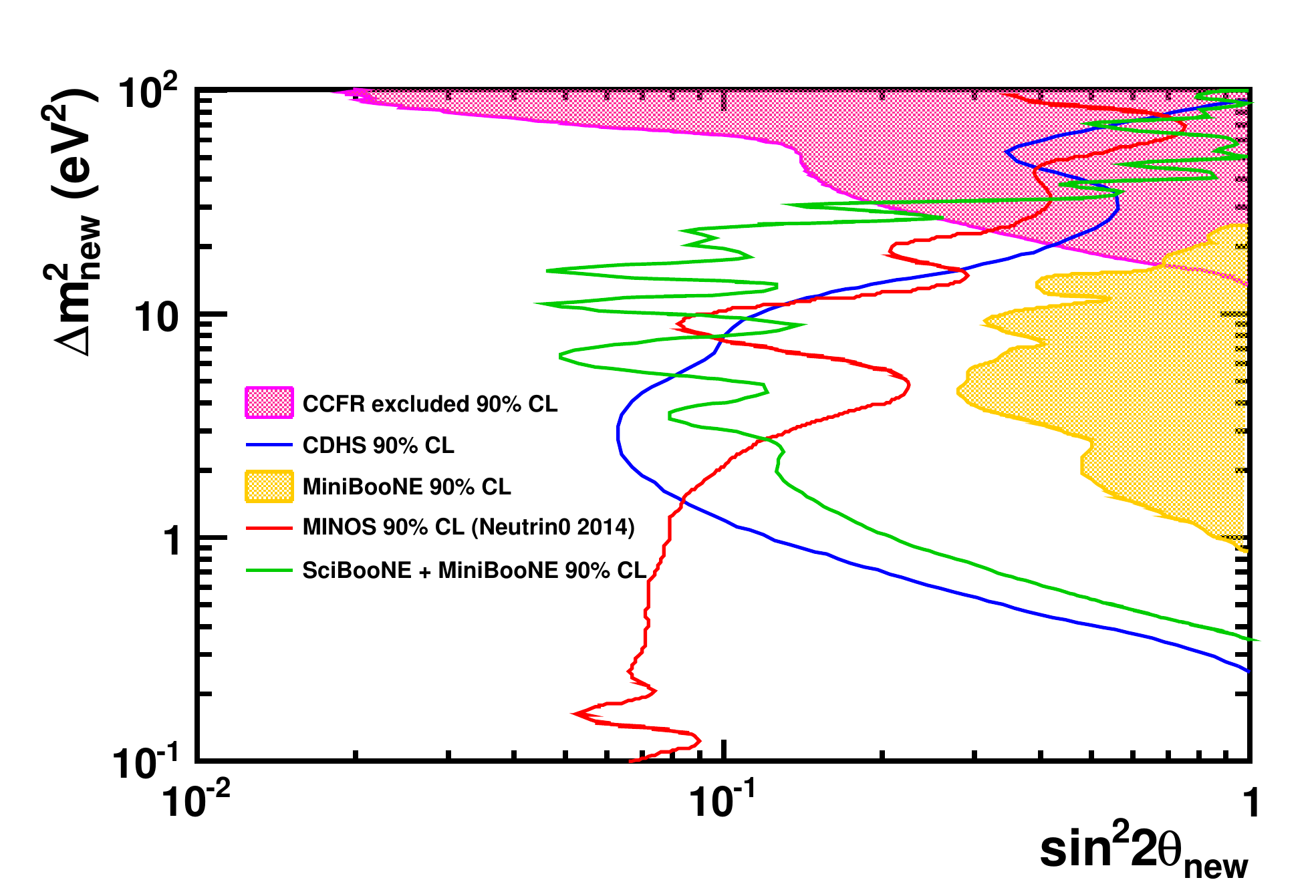}%
\caption{\label{numu_disapp}The current exclusion limits from $\nu_\mu$ disappearance searches in the 0.1 to 100 eV$^2$ $\Delta m^2$ region with the mixing angle given by $\sin^2 2\theta_{new}$.
 All the exclusion limits are at 90\% C.L. (From Ref.~\cite{numu_dis}).}
\end{figure}

The $\nu_\mu$ disappearance experiments are looking for a deficit in detected muon neutrino events as compared to the predicted number of events.  The prediction is best done by using an experimental setup with two detectors, a near detector at short distance to measure the flux and a far detector to search for the deficit. Extrapolating from the near to far detector can bring in systematic uncertainties that depend on modeling the beam flux and divergence as well as detector differences but these uncertainties are typically at the few percent level.  
  Fig.~\ref{numu_disapp} shows the current limits for the various measurements.  CCFR, CDHS, MINOS, and SciBooNE/MiniBooNE are all two detector measurements and the MiniBooNE only limit comes from a shape analysis of the observed events compared to expectation. 

\begin{figure}[t]
\centering
\includegraphics[width=0.65\linewidth]{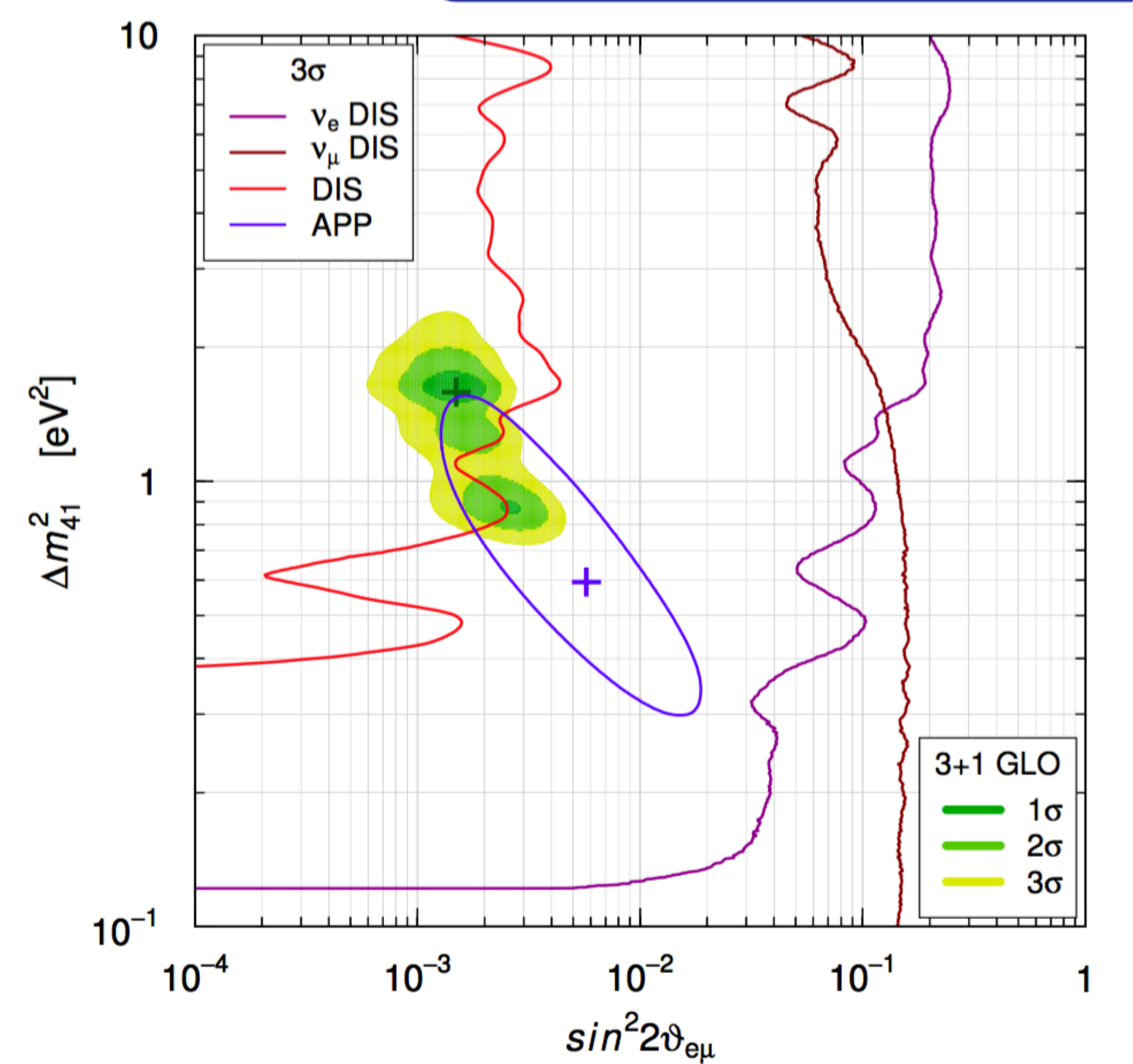}
\caption{ \label{Giunti_GLO}
Allowed regions (at 3$\sigma$) in the $\Delta{m}^{2}_{41}$ versus $\sin^22\theta_{e\mu}$ plane from global fits of short-baseline appearance and disappearance oscillation data (3+1 GLO) compared with the allowed regions obtained from  $\nu_{\mu}\rightarrow\nu_{e}$ 
appearance-only data (APP)
and the constraints obtained from 
$\nu_{e}$
disappearance-only data ($\nu_{e}$ DIS), 
$\nu_{\mu}$
disappearance-only data ($\nu_{\mu}$ DIS) and the
combined disappearance data (DIS).
The best-fit points of the GLO and APP fits are indicated by crosses.
(Updated plot presented by C. Giunti at the Neutrino 2016 conference associated with Ref.~\cite{Giunti2016_GLO}.)
}
\end{figure}

The tension between the appearance and disappearance data is displayed in Fig.~\ref{Giunti_GLO} which shows the oscillation parameters associated with different data sets at the 3$\sigma$ CL. As given in Eq. \ref{3+1relat} ($\sin^22\theta_{\mu e} \approx \frac{1}{4} \sin^22\theta_{ee}\sin^22\theta_{\mu\mu}$), the independent $\nu_\mu$ and $\nu_e$ disappearance limits can be used to give a 
combined DIS limit on $\sin^22\theta_{\mu e}$ for 3+1 models.  
One can see that the combined $\nu_e$ allowed region (APP) is highly restricted by the DIS limits.   One should keep in mind that limits are not hard cutoffs, and there is lower probability signal extending outside of the defined 3$\sigma$ CL, where APP and DIS do overlap.
A global fit of all data is also shown and indicates the small allowed region left from the APP allowed region.   

\begin{figure}[t]
  \centering
\includegraphics[width=0.75\textwidth]{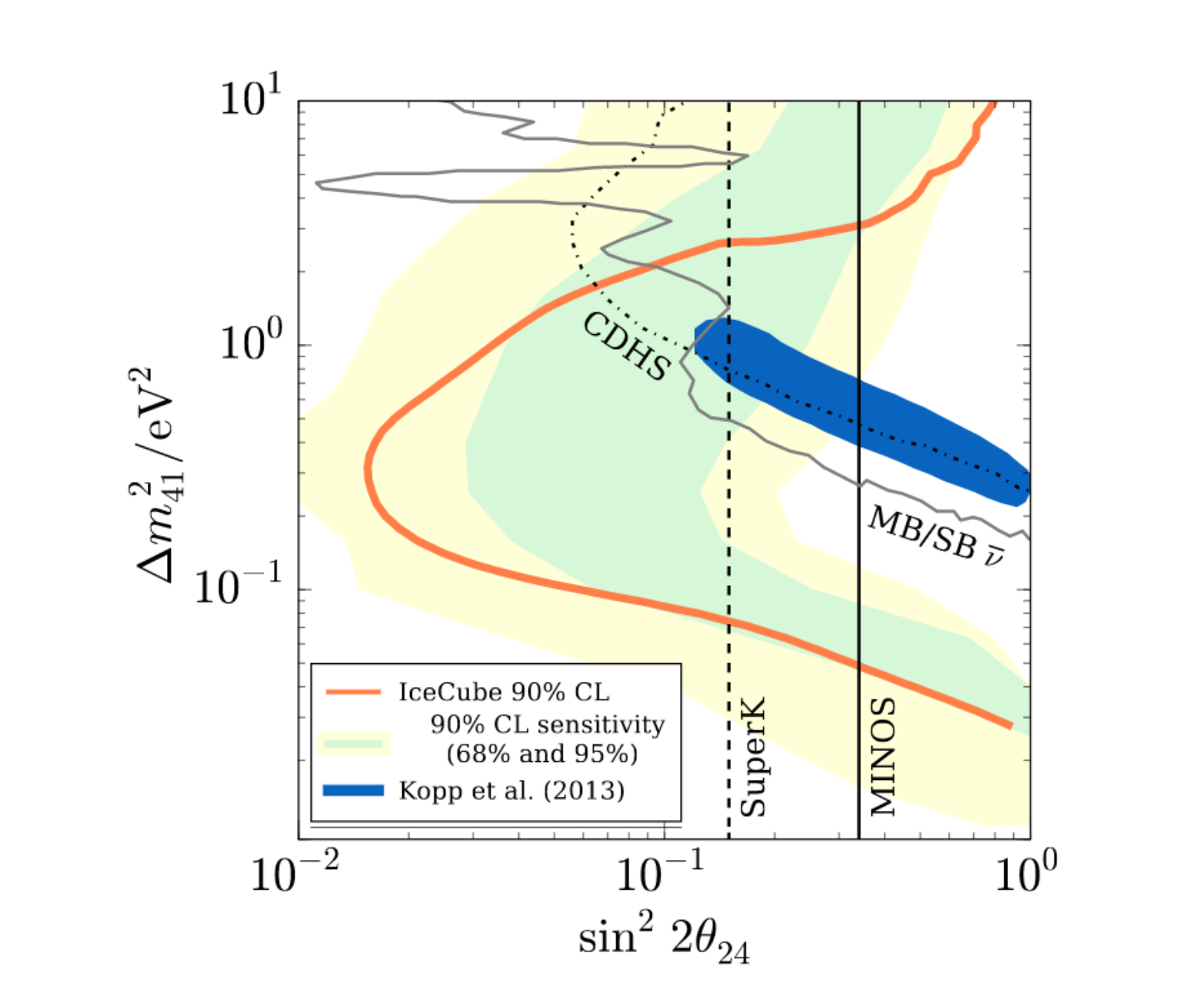}%
\caption{\label{IceCube}Results from the IceCube $\nu_\mu$/$\bar\nu_\mu$ disappearance search.  The 90\% (orange solid line) CL contour is shown with bands containing 68\% (green) and 95\% (yellow) of the 90\% contours in simulated pseudo-experiments. Also, shown are the CDHS and SciBooNE/MiniBooNE limits along with the  MiniBooNE/LSND 90\% CL blue allowed region \cite{Kopp_global} assuming $|U_{e4}|^2= 0.023$. (From Ref.~\cite{IceCube}).}
\end{figure}

The IceCube experiment has also done a search for $\nu_\mu$ and $\bar\nu_\mu$ disappearance using the atmospheric muon neutrino spectrum as a function of zenith angle as described in Sec.~\ref{matter}.  As a reminder, this experiment depends on the modification of the vacuum oscillations by matter effects.    As shown in Fig.~\ref{IceCube}, these results greatly improve the limits from previous experiments for $\Delta m^2 < 2$ eV$^2$.    
Note that this figure shows the LSND and MiniBooNE allowed regions, which are for appearance, on the disappearance plot.  The appearance allowed regions were transferred to this plot using the best fit value for $|U_{e4}|$ rather than a point-by-point calculation of $|U_{e4}|$ in a global fit involving IceCube.   As was pointed out in Sec.~\ref{LE}, this will lead to some distortion of the appearance allowed region.   So the overlay should be regarded as qualitative, not quantitative.
Nevertheless, this result only adds the the apparent tension between appearance and disappearance data sets.

In summary, the current sterile neutrino oscillation status has several signals in the $\nu_e$ appearance channel, and also signals in the $\nu_e$ disappearance channel, but no oscillation signal for $\nu_\mu$ disappearance.    A smoking gun for oscillations would be the observation of detected event rates that vary with the expected $L/E$ behavior; currently, this behavior has not been convincingly observed in any of the experiments with $\Delta m^2\sim 1$ eV$^2$ signals.

\section{Global Data Fits  \label{globfits}}

In order to determine the viability of sterile neutrino models,  it is necessary to use a global fit.   This needs to include as many of the relevant data sets as possible.   It is best if these data sets are from considerably different experimental designs, as this reduces sensitivity to backgrounds that are under or oversubtracted and to other systematic uncertainties.     A few experiments have recently adopted the practice of presenting their limit combined with one other experiment that will give the best reach.   This turns out to be misleading and the results over-predict the final result of the global fit.  This should be avoided.

In this section we discuss the techniques of global fits.    An important aspect of the fit is the choice of test statistic. This turns out to be a complicated issue, which we must consider first.     We then discuss implementation.   Throughout, we then show results of global fits to the present data,  however our emphasis is on the approach to the problem rather than these specific results, as these change rapidly with time.

\subsection{Test Statistics for Global Fits}

\subsubsection{The problem with the $\chi^2$ statistic for fitting \label{chi2_fit}}

The simplest test statistic for a fit is the $\chi^2$ variable, which compares the data to the prediction including uncertainties.  The distribution of this variable for Gaussian uncertainties should have a mean equal to the $dof$ and standard deviation equal to $\sqrt{2\ dof}$.  However, this statistic can be very misleading because the global fits are keying in on deviations of the data from the prediction in rather localized data regions that are sensitive to oscillations.   For the data space outside of these sensitive region, one can get a good fit to almost any oscillation model.    If you add these insensitive bins into the $\chi^2$, then the power to discriminate models will be diluted and $\chi^2/dof \rightarrow 1$.    

\begin{figure}[t]
\vspace{+0.1in}
\centerline{\includegraphics[angle=0, width=1.0\textwidth]{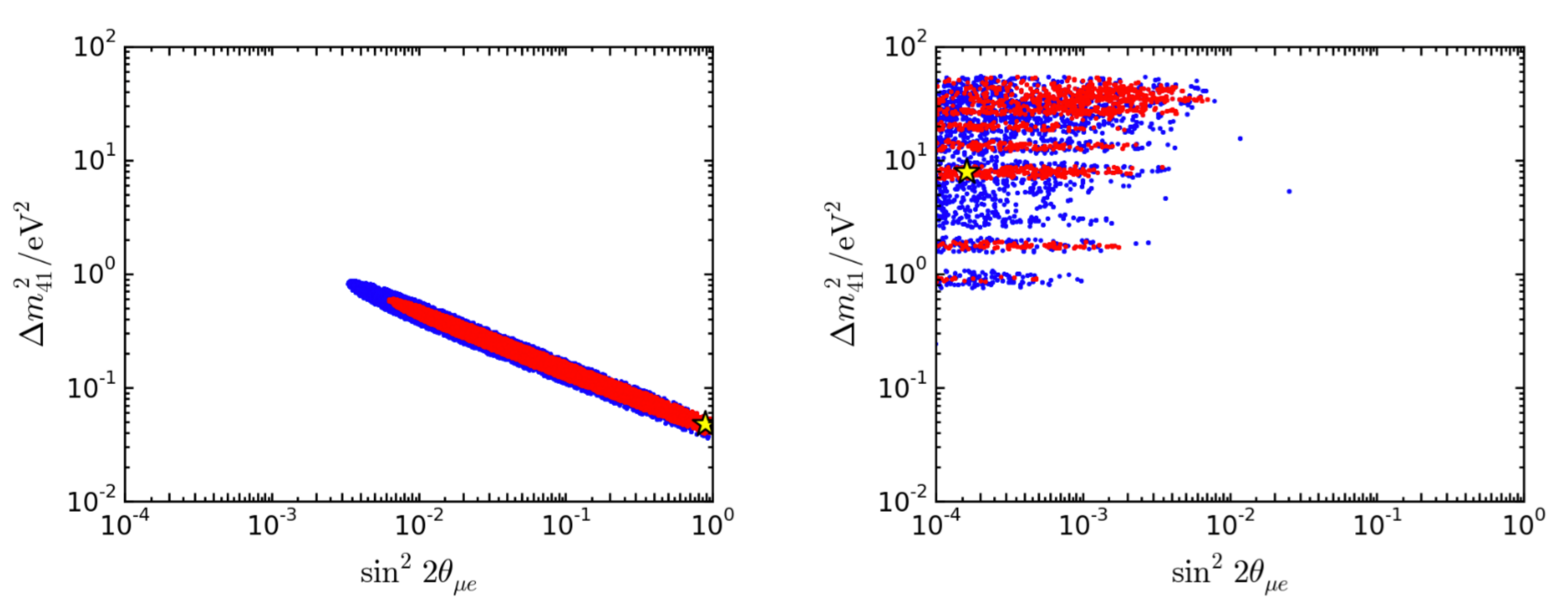}}
\vspace{-0.1in}
\caption{Left:  global fit to the appearance data only;  Right: global fit to the disappearance data only. Plot is from Ref.~\cite{SBL2016}} 
\label{appdiscompare}
\vspace{-0.1in}
\end{figure}

This problem is enhanced when one is fitting many data sets, which all may have insensitive bins in certain areas of parameter space that dilute the $\chi^2$.   As a result, even though data sets may disagree,   the $\chi^2/dof$ from the
global fit may be acceptable.   

This is the case with the present global fits.   In a 3+1 model,  the $\chi^2/dof = 359/315~dof = 1.14$ for all of the data.  However, separate fits to the appearance and disappearance data sets listed on Table~\ref{tab:explist} sets show poor overlap as seen in Fig.~\ref{appdiscompare}.   One ends up with a global fit result with allowed regions because the strength of the combined $\nu_e$ disappearance and $\nu_\mu \rightarrow \nu_e$ appearance results overcomes the lack of signal in $\nu_\mu$ disappearance. But the $\chi^2/dof$ does not reflect the stress due to the large number of bins in all three cases where the best fit model predicts no signal.

One does not want to change the selected binning to focus on only the identified signal regions as one performs a fit.  This would make a test statistic that could only be understood with extensive frequentist tests using a large set of experiment simulations that, at present, are computationally too expensive to perform.     Also, the regions with no signal can add very important information to the fit and be used, for example, to constrain normalization uncertainties.

\subsubsection{The $\Delta \chi^2$ statistic for global fitting}

\begin{figure}[t]
\vspace{+0.1in}
\centerline{\includegraphics[angle=0, width=0.6\textwidth]{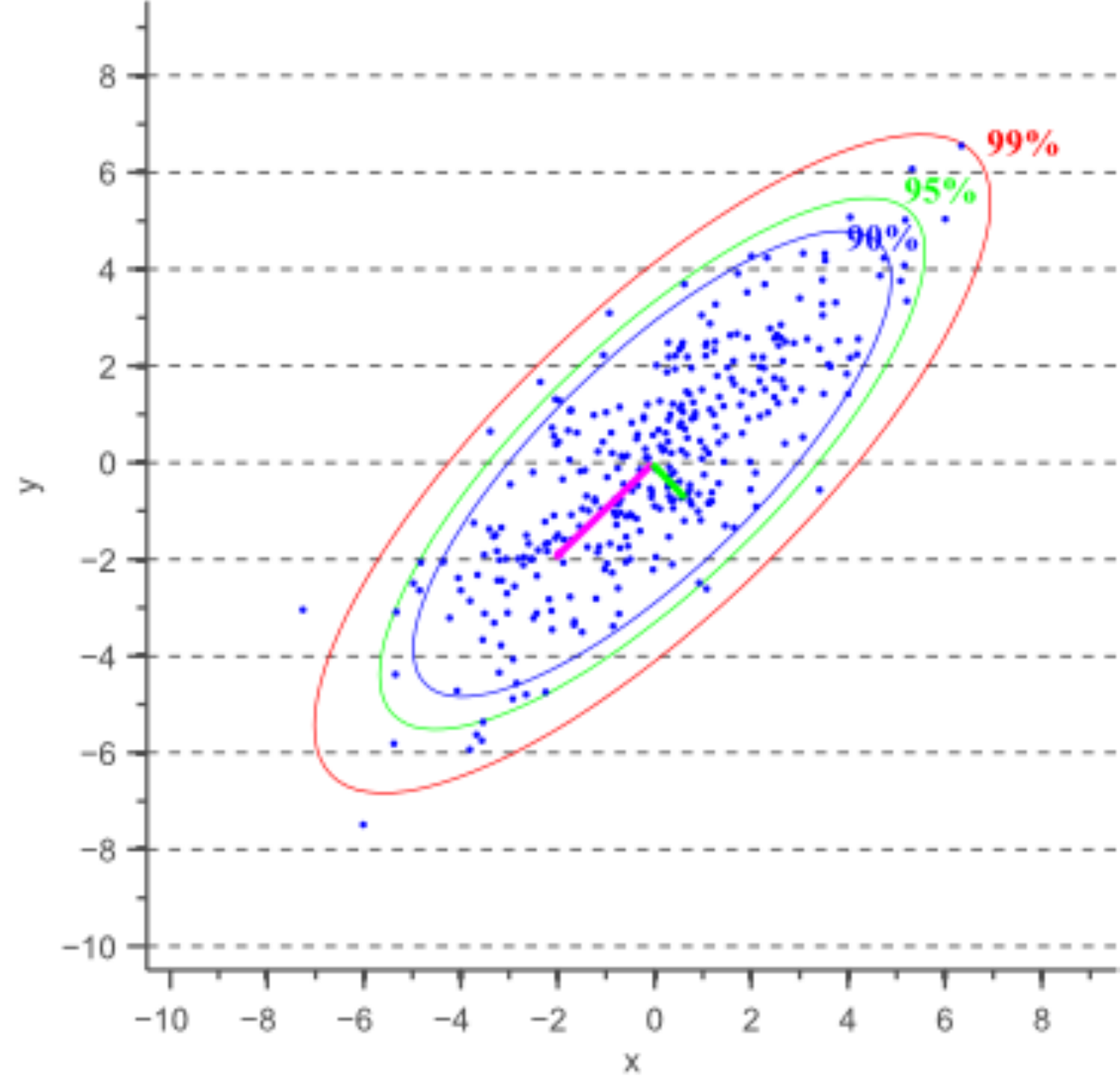}}
\vspace{-0.1in}
\caption{Example confidence level (CL) regions for a two-dimensional, normally-distributed data sample in the variables $(x,y)$. The arrows show the variance of the data along the two covariance (or eigenvector) directions. The ellipses show the CL regions at 90\%, 95\%, and 99\% that contain that fraction of the data points. (From Ref.~\cite{2D_cartoon})} 
\label{CL_cartoon}
\vspace{-0.1in}
\end{figure}

A powerful method for doing global fitting to determine oscillation parameters and confidence regions is to use the $\Delta \chi^2$ statistic, where 
\begin{equation}\label{Delta_chi2}
  \Delta \chi^2 = \chi^2_{TestPoint} - \chi^2_{BestFit}.
\end{equation}
The $\Delta \chi^2$ statistic is related to the likelihood $(L)$ ratio of the best fit with respect to some test point for Gaussian uncertainties with 
\begin{equation}
\Delta \chi^2 = -2\ln{(L_{TestPoint}/L_{BestFit})}.
\end{equation}
The best fit for a given data set is found by minimizing the $\chi^2$ for the data versus prediction.  For the simplest problems, $\Delta \chi^2$ will follow a $\chi^2$ distribution with the $dof$ value equal to the number of parameters being determined in the fit.  To determine a CL region around the best fit, one uses the critical value for the $\chi^2$ distribution with this given $dof$.  For example with two fit parameters (say $\Delta m^2$ and $\sin^22\theta$), the critical value for 90\% CL would be 4.61 and for a 2$\sigma$~CL would be 6.18.  One then can find the $\Delta m^2$ {\it vs.} $\sin^22\theta$ region where the $\Delta \chi^2$ value for those points are below this critical value.  This region then corresponds to the allowed region at the given CL.
An example of CL regions for a toy model with two-dimensional, normally-distributed data is shown in Fig.~\ref{CL_cartoon}.

Limits are a special case of confidence intervals where, for example, the no-oscillation (or ``null'') model is allowed by the data.  On a $\Delta m^2$ {\it vs.} $\sin^22\theta$ plot, a limit curve, at a given CL, separates the allowed region to left of the curve from the excluded region to the right of the curve.  The curve corresponds to the points where $\Delta \chi^2$ is equal to the critical value for the given CL.  Thus, the points to the right of the curve have $\Delta \chi^2$ values greater than the critical value and are, therefore, excluded at the given CL.

Other parameters that are not directly involved in the oscillation model ({\it a.k.a.} ``nuisance'' parameters) such as those that describe the normalization, backgrounds, etc. and their uncertainties, can be included in the fit using extra terms in the $\chi^2$, called ``pull'' terms.  These nuisance parameters can be included (usually called being ``profiled'') in the fit by minimizing them at each test point.  The fit value for the nuisance parameters can be of interest and comparisons of the values of these parameters with their expectations can give information on how much the fit determines they have been pulled from what is expected.  This is commonly referred to as the ``pull'' for a parameter.  

The $\Delta \chi^2$ also solves the problem described in Sec.~\ref{chi2_fit} where data bins outside of the sensitive region erroneously reduce the $\chi^2/dof$.  For the $\Delta \chi^2$ calculation, bins with no or very small sensitivity in the fit will contribute zero to the $\Delta \chi^2$ value since the contribution from the test point and from the best fit will be identical and cancel.

A problem with the $\Delta \chi^2$ statistic is that the probability distribution may not follow a $\chi^2$ distribution with a $dof$ equal to the number of fit parameters.  This can come about because the predicted event function is non-linear in the fit parameters or because the uncertainty in certain bins is hard to determine from first principles or is non-Gaussian.  An example of this behavior is given by fits in the high $\Delta m^2$ region,  much above the value for the first oscillation maximum.  In that region, the $L/E$ behavior in the oscillation probability has rapid oscillation and effectively gives an average of 0.5 for the $\sin^2(1.27 \Delta m^2 L/E)$ factor.  Therefore, in this region the oscillation probability has no dependence on $\Delta m^2$ and the effective $dof$ number will be 1.0 instead of 2.0.  To quantitatively calculate the $dof$ value as a function of test point parameters and, thus, the $\Delta \chi^2$ critical values for a given CL, one needs to do simulation studies.  At every test point, one uses a large set of fake, simulated experiments to determine the $\Delta \chi^2$ distribution numerically from the distribution of simulated values.  This is commonly referred to as the ``Feldman-Cousins'' frequentist method \cite{Feldman_Cousins} for determining confidence regions.

\subsubsection{Anscombe's Quartet as a Cautionary Tale}

In Sec.~\ref{currentresults}, we emphasized that there is tension between the appearance and disappearance data sets.   As we will present below, the $\Delta \chi^2$ test statistic will lead to apparently high-quality fits, despite this tension.     It is reasonable to ask how this happens.   The example of Anscombe's Quartet \cite{Anscombe} provides an explanation of how this can come about.

\begin{figure}[t]
\vspace{+0.1in}
\centerline{\includegraphics[angle=0, width=0.6\textwidth]{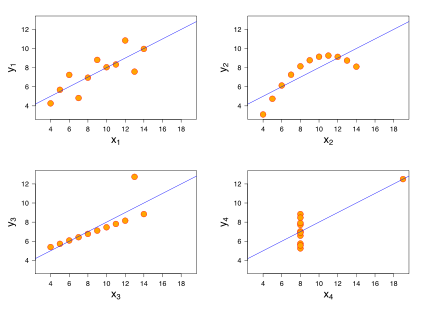}}
\vspace{-0.1in}
\caption{Anscombe's quartet of data sets (From Ref.~\cite{wikiAnscombe} and first published in Ref.~\cite{Anscombe}).  This shows four data sets, that all fit to the same straight line.    The data sets have the same mean and variance.} 
\label{Anscombefig}
\vspace{-0.1in}
\end{figure}

Figure~\ref{Anscombefig}, reproduced from ref.~\cite{wikiAnscombe}, shows four example distributions.   All four distributions are fit to lines, yielding slopes and intercepts that agree to three and two digits past the decimal, respectively.    They also have the same mean, variance, total $\chi^2$,  and correlation between $x$ and $y$.   The information from a fit to all data points does not indicate that these are substantially different distributions.   However, the differences between these distributions becomes clear if you divide the data sets in certain ways and perform separate fits on each subset.   In that case, the fits to the two separate subsets will not agree with each other, nor with the fit to all of the data points.

Having discovered a problem like this, the question then becomes: how do you quantify the probability of the split between two data sets resulting in the inconsistent?    The problem is not simple--by selecting how you will cut the data set, you no longer have random samples.   You have biased the information by choosing how to cut the data set to maximize the tension.   A frequentist fake-data study will tell you how often you will see such a disagreement or worse disagreement.  But such a study is prohibitive in the case of the short baseline global fits, given today's computing power.      

This problem led to the introduction of the Parameter Goodness of Fit, which we describe below.     It provides a useful approach to comparing data sets that you have split.   However, we will show that this approach fails in certain circumstances.    At this point, the meaning of the probability returned by the Parameter Goodness of Fit remains unclear.

\subsubsection{The Parameter Goodness of Fit}

A possible problem with the $\Delta \chi^2$ global fit method to determine confidence regions is that there can be large discrepancies between classes of data sets that go into the fit.  This may lead to allowed regions deriving from the $\Delta \chi^2$ prescription, even when no part of the parameter space can well-reproduce the data. An example of this behavior is shown in Fig.~\ref{appdiscompare}, where the fits to appearance and disappearance give much different allowed regions but taken together do have allowed regions at reasonable confidence levels. Of course, the global fits should include all the uncertainties in determining the allowed region.  Nevertheless, one might also want to quantify how compatible are the different data sets that go into this determination. 

In response to this question,  the Parameter Goodness-of-fit (PG)
test was established \cite{pgtest}.    This test was meant to provide a
compatibility test for two subsets of data in a given global fit,
without bias that may come from irrelevant bins or data sets with many
bins.     Ref.~\cite{pgtest} provides a formal, analytic derivation of
this new test statistic, the $\chi^2_{PG}$, its degrees of freedom,
$N_{PG}$ and the cumulative
probability function for this PG test statistic.

As an example of its application,  consider the PG test for
appearance ($\nu_\mu \rightarrow \nu_e$)
versus disappearance ($\nu_\mu \rightarrow \nu_\mu $ and $\nu_e \rightarrow \nu_e$) data sets that underlie a 3+1 global fit.
One performs separate fits on each of the two underlying subsets,
as well as a combined fit to the full data set, to obtain three $\chi^2$ values,
$\chi^2_{app}$, $\chi^2_{dis}$ and $\chi^2_{glob}$.   One then
forms an effective $\chi^2$:
\begin{equation}
\chi^2_{PG}=\chi^2_{glob}-(\chi^2_{app}+\chi^2_{dis}).  \label{chi2pg}
\end{equation}  
In the case of good agreement between underlying data sets,   the $\chi^2$ contribution
from non-signal bins is highly reduced because of the subtraction in a similar way as the $\Delta \chi^2$ statistic.  The number of degrees of
freedom is defined as 
\begin{equation}
N_{PG}=(N_{app}+N_{dis}) - N_{glob}~   \label{npg}  
\end{equation}
where each $N$ is the number of independent parameters involved in the
given fit.      In our 3+1 example,   the appearance fit has two
parameters ($|U_{e4}||U_{\mu4}|$ and $\Delta m^2$), the disappearance fit has three parameters ($|U_{e4}|$, $|U_{\mu4}|$, and $\Delta m^2$), and the global
fit has three parameters ($|U_{e4}|$, $|U_{\mu4}|$, and $\Delta m^2$).  This then leads to the $dof$ calculation for the PG test of   $N_{PG}^{3+1} = (2+3)-3=2$ degrees of freedom.
The probability that the two data sets are in agreement is
that associated with $\chi^2_{PG}$ for  $N_{PG}$, or $\chi^2_{PG}$ for $2~dof$ in the
case of a 3+1 model.

The PG test has been shown to successfully assess the agreement
between data sets with systematic uncertainties 
that are Gaussian-distributed across multiple runs of the experiment.  
However, in the case of  an underlying systematic which
is single-valued rather than 
Gaussian-distributed between runs of an experiment, the
number of degrees of freedom is not necessarily $N_{PG}$.   
The deviation occurs when the shape of the systematic effect is 
correlated to the shape of the oscillation signal.
In this case,  $\chi^2_{PG}$ for $N_{PG}$ is not a valid estimate of the probability.

As a tangible example, imagine an appearance experiment which has
both a 3+1 oscillation signal and a known background.     To fit the
data, the background must be subtracted.   For simplicity,  let's 
assume the background shape is perfectly known from the Standard Model, but the normalization
has an error that comes from past measurements of the cross section.      Our imaginary 
experimentalists will look up the central value for the background
assigned by the Particle Data Group \cite{pdg}, and use this to 
subtract a background function with this normalization.     The
assumption is the data set now has zero background, after subtraction,
but that there is an associated error that comes
from the past measurements of the cross section.    This error represents the
experimentalist's best knowledge of the cross section, and it is 
presented and treated as a Gaussian-distributed systematic error.      In reality, however,
the true, natural value of the background is single valued.    No
matter how many times the experiment is run,  this true value is always the
same from run to run.   Therefore,  the experiment has an underlying
residual background subtraction which is the same for every run and is not Gaussian-distributed.      

To make our imaginary experiment more concrete, consider the following legitimate set of oscillation 
parameters for a 3+1 model:   
\begin{itemize}
\item $\Delta m^2=0.75$ eV$^2$,  
\item $|U_{e4}|^2=0.1$, and 
\item $|U_{\mu
  4}|^2=0.1$.   
\end{itemize}
Thus $|U_{e4}||U_{\mu 4}|=0.1$.
These are used to generate the ``true'' oscillation signals that our imaginary experiment will see.   We generate data for $\nu_\mu
\rightarrow \nu_e $ appearance, $\nu_\mu$ disappearance and $\nu_e$
disappearance in 16 energy bins in the range from 200 to 1000 MeV,
assuming $L=500$ m.

The problem occurs when the residual background for our imaginary experiment
has the same shape as a legitimate oscillation signal, but one with different parameters than the model we describe above.   As an example, we consider a background to the appearance signal that is an 
exponential residual background of the form $N^{backgnd} = A_{true}\exp(-E/200.)$,
  where $E$ is the neutrino energy.
We will study values of the normalization of the residual background,
$A_{true}$, between 0.0 and 0.4.  

(We leave it as an exercise for the student to explore this model with zero background and with a 
flat residual background as a function of neutrino energy of the form $N^{backgnd}_i=A_{true}=constant$.   
Note that the latter case is important because a 3+3 model with large $\Delta m^2_{61}$ can lead to a flat overall offset.
In these two cases, the student will find the PG test succeeds.) 

Our imaginary experimenters have assumed that this background has been correctly subtracted,  thus
$A_{exp}=0$.     
The shape is known and so they 
fit for the normalization, $A_{fit}$.
The experimenters places a systematic error on
this assumption, which, for the purposes of this discussion, will be 
$\sigma_{A_{exp}}=0.15$.

In order to generate the appearance data,  the
function describing the residual background is added to the function
describing the oscillation signal,  resulting in a function that
describes the total $\nu_e$-like events.      The same oscillation
parameters and residual background function are used to generate each
data set in a study.

The experimental data is then generated using statistical uncertainties about
the functions for the appearance (with background) and disappearance predictions.
This simulates statistical fluctuations but no other smearing effects are included.
We will study the effect of the underlying background by running 1000
fake studies, each of which has an appearance, electron disappearance
and muon disappearance experimental data set.

In each of the 1000 fake studies,  $A_{true}$ is held to the same
value.   Each data set is fit for the oscillation parameters with the
appearance and global data set also simultaneously fit for the
normalization of the residual
background, $A_{fit}$, using a pull term. 
We define the $\chi^2$ for each fit in the follow way: 
\begin{eqnarray}
 \chi _{\nu _e app}^2  &=& \sum\limits_{i = 1}^{16} {\frac{{\left(
         {d_i^{\nu _e app}  - \left( {osc_i^{\nu _e app}  + b_i^{\nu
                 _e app} \left( {A_{fit} } \right)} \right)} \right)^2
     }}{{\left( {\sigma _i^{\nu _e app} } \right)^2 }}}  \nonumber \\
&&~~~~~~~+
 \frac{{\left( {A_{fit}  - A_{exp} } \right)^2 }}{{\sigma
     _{A_{exp} }^2 }}  \nonumber \\  
      \label{appeq} 
   \end{eqnarray}  
   
\begin{eqnarray}
 \chi _{disapp}^2 & = &\sum\limits_{i = 1}^{16} {\frac{{\left(
         {d_i^{\nu _\mu  disapp}  - osc_i^{\nu _\mu  disapp} }
       \right)^2 }}{{\left( {\sigma _i^{\nu _\mu  disapp} } \right)^2
     }}}  \nonumber \\
&&+ \sum\limits_{i = 1}^{16} {\frac{{\left( {d_i^{\nu _e disapp}  - osc_i^{\nu _e disapp} } \right)^2 }}{{\left( {\sigma _i^{\nu _e disapp} } \right)^2 }}} \nonumber \\
\label{diseq}
   \end{eqnarray}  
   
\begin{eqnarray}
 \chi _{global}^2 &= &\sum\limits_{i = 1}^{16} {\frac{{\left( {d_i^{\nu
             _e app}  - \left( {osc_i^{\nu _e app}  + b_i^{\nu _e app}
               \left( {A_{fit} } \right)} \right)} \right)^2
     }}{{\left( {\sigma _i^{\nu _e app} } \right)^2 }}}  \nonumber \\
&&~~~~~~~+
 \frac{{\left( {A_{fit}  - A_{exp} } \right)^2 }}{{\sigma
     _{A_{exp} }^2 }}  \nonumber \\
 &&+ \sum\limits_{i = 1}^{16} {\frac{{\left( {d_i^{\nu _\mu  disapp}
           - osc_i^{\nu _\mu  disapp} } \right)^2 }}{{\left( {\sigma
           _i^{\nu _\mu  disapp} } \right)^2 }}}  \nonumber \\ 
&&+ \sum\limits_{i = 1}^{16} {\frac{{\left( {d_i^{\nu _e disapp}  - osc_i^{\nu _e disapp} } \right)^2 }}{{\left( {\sigma _i^{\nu _e disapp} } \right)^2 }}}.  \label{globeq}
 \end{eqnarray}

\begin{figure}[t]
\begin{center}
\includegraphics[width=0.75\textwidth]{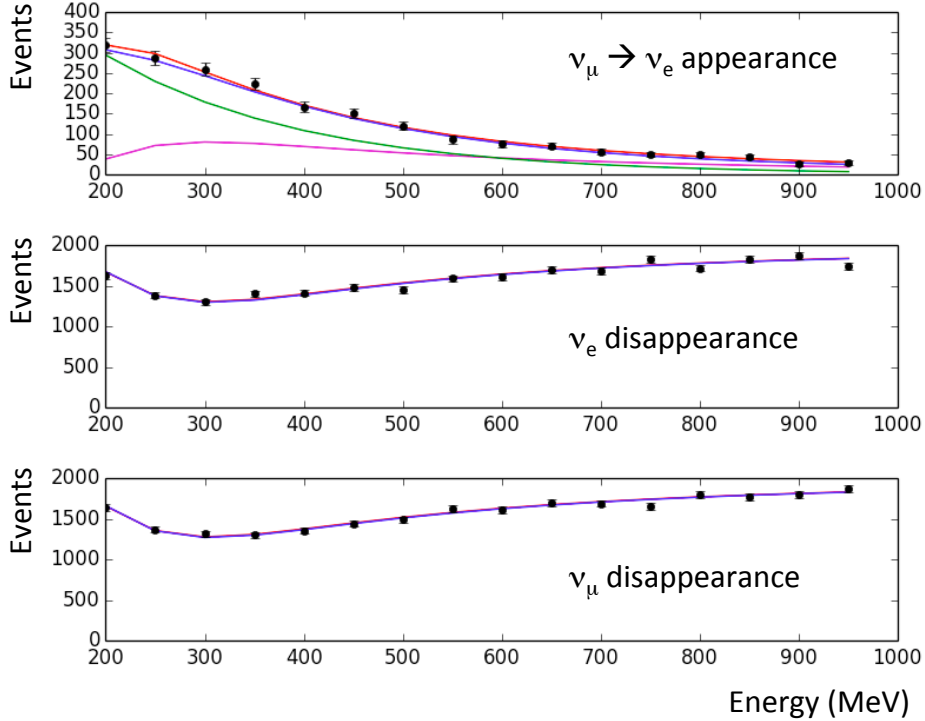}
\end{center}
\vspace{-0.25in}
\caption{\it \footnotesize Example of signals versus energy for appearance (top),  electron-flavor disappearance
  (middle) and muon flavor disappearance (bottom) from one experiment
  with an exponential residual background of $A_{true}=0.4$.  In all
  plots, the true signal is shown in magenta.    In the top
  plot, the underlying background is shown in green.   In all plots,
  the global fit result is shown in blue.    In the top plot, the
  appearance-only fit, with the residual amplitude included as a pull
  term, is shown in red.
\label{generatedexp} }
\end{figure}

Figure \ref{generatedexp} provides an example run for the case of a
residual exponential background where $A_{true}=0.4$.     The top,
middle and bottom frames, are, respectively, appearance,
electron-flavor disappearance, and muon-flavor disappearance signals
shown as a function of energy.   The points show the generated data
with statistical fluctuations.     The true oscillation signal is shown in
each frame in magenta.    The green curve in the appearance data set
(top frame) shows the underlying residual background for this run.
The red curve on the appearance data set, shows the fit obtained from
minimizing  $\chi _{\nu _e app}^2$ (Eq.~\ref{appeq}).    The blue
lines on all three data sets show the fits obtained by minimizing
$\chi _{global}^2$ (Eq.~\ref{globeq}).   

\begin{figure}[t]
\begin{center}
{\includegraphics[width=1.0\textwidth]{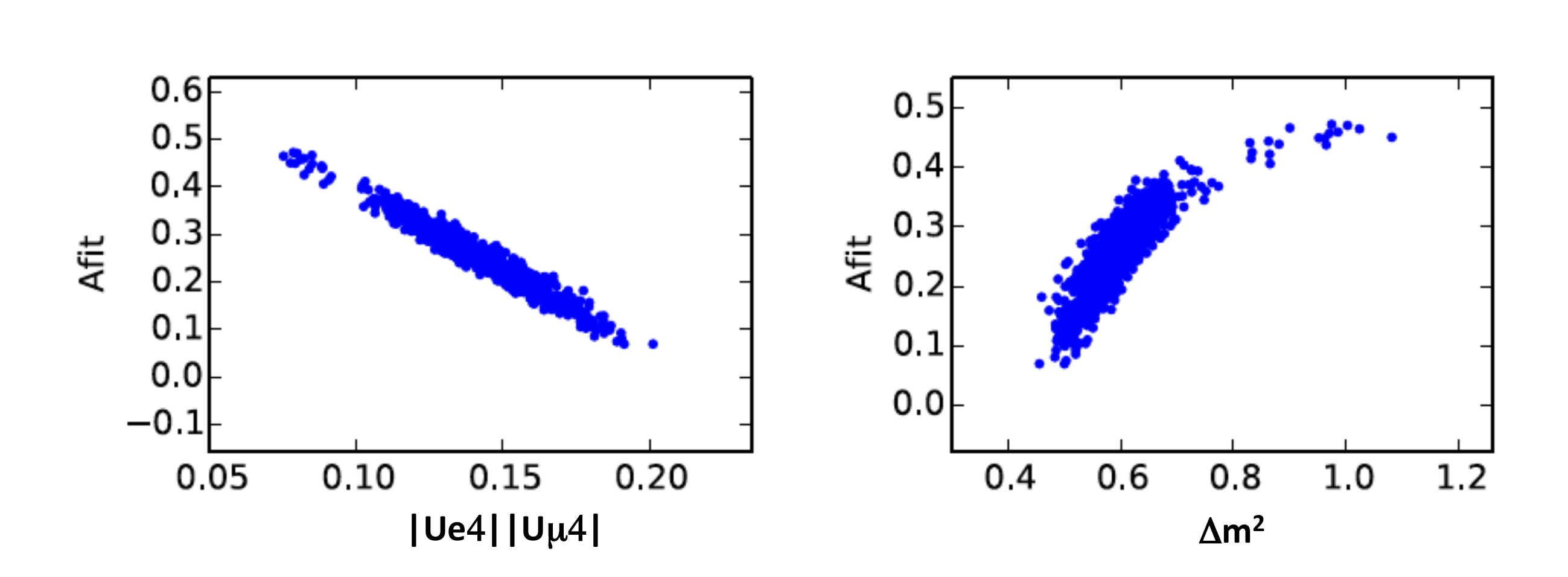}}
\end{center}
\caption{\it   Parameter values for the appearance fit for an
  exponential background  with $A_{true}=0.4$.  One can see strong
  correlations between the parameters, and the fit values deviate far from
  the true values.   
\label{correlations}}
\end{figure}

In this case of an exponential background, the shape of the
oscillation signal and the background are correlated with a correlation parameter of $\rho=0.6$.   
This correlation is reflected in the range of values that the
parameters can take in any given run, as is illustrated in
Fig.~\ref{correlations}.    These plots are for the case of $A=0.4$,
and the plots shown are for the appearance fit.    One can see that
the parameters are widely varying from the true value and they are 
highly correlated.    

On the other hand,  the global fit performs very well.     The
oscillation parameters that are returned are consistently in agreement
with the input values.  We find 
 $A_{fit} =0.100\pm0.011, \/\/\/\/0.0198\pm0.013, \/\/\/\/0.297 \pm 0.016, \/\/\/\/0.395\pm0.019
$ for $A_{true}$=0.1, 0.2, 0.3, and 0.4, respectively.

In order to find the 
degrees of freedom,  the $\chi^2$ distributions of the data sets are  
fit to a standard $\chi^2$ form  
\begin{equation}
P_0 ( x^{(P_1/2)-1)} e^{-x/2}. \label{dofeq}
\end{equation}
where $x=\chi^2$, $P_0$ is a normalization and $P_1$ is the number of degrees of freedom indicated by the $\chi^2$ distribution.

\begin{figure}[t]
\begin{center}
{\includegraphics[width=0.75\textwidth]{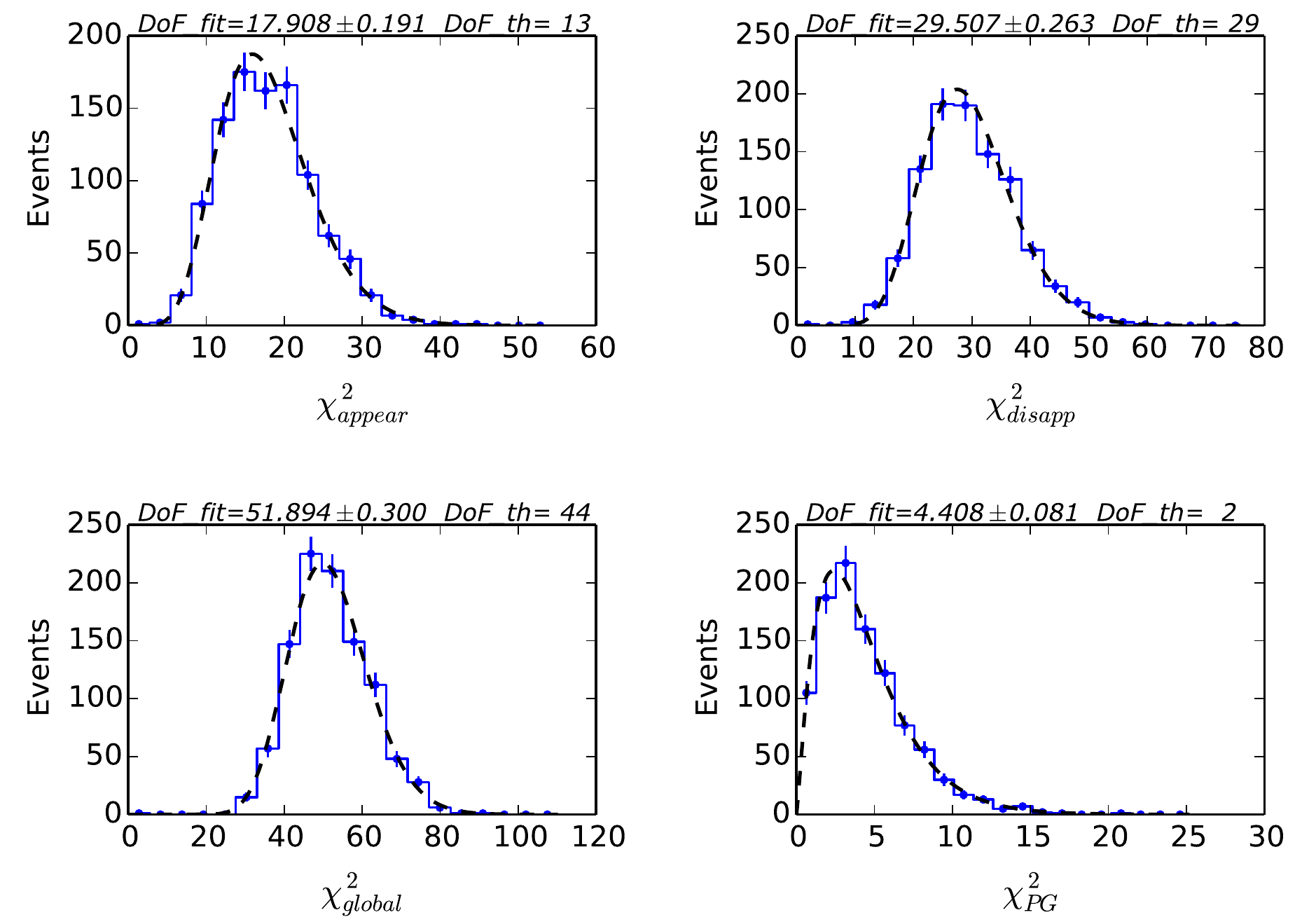}}
\end{center}
\caption{\it   The $\chi^2$ distributions for an exponential background with
  $A_{true}=0.4$. For each of the $\chi^2$ types, the results of the $dof$ fit are given with uncertainties along with the theoretical expectation. The histogram with error bars are the distribution of values and the dashed curve is the result of the $dof$ fit.  The number of degrees of freedom from the fit for the $\chi^2_{PG}$ deviates significantly from the expected value of two.
\label{expfitdof}}
\end{figure}

As an example of the $dof$ fits, the $\chi^2$ distributions for 1000 experiments for the case of
$A_{true}=0.4$ are shown in Fig.~\ref{expfitdof}.   One can see that
in this case,  the $\chi^2_{PG}$ deviates far from the expectation of
two even though there is not an incompatibility of data with a true oscillation model encompassing all three data sets.
For $A_{true}$=0.1, 0.2, 0.3, and 0.4, respectively, the degrees
of freedom of the $\chi^2_{PG}$ are found to be:  $1.92\pm 0.062$, 
$2.484\pm 0.065$, $3.361\pm0.074$ and $4.408\pm 0.081$.  
The problem is arising because the high correlation between the
background and the signal allows the fit to find a minimum
$\chi _{\nu _e app}^2$ whose distribution is close to the theoretical expectation, even though the
fit parameters are far from the inputs.  On the other hand, the global fit
parameters are constrained by the disappearance data contributions and so the minimum
value of the $\chi^2_{global}$ is relatively large.    This leads to
a $\chi^2_{PG}$ distribution that is larger than the expectation for 2 $dof$.  Since the $\chi^2_{PG}$ values are larger than what is theoretically expected, the PG test will on average erroneously indicate that the appearance and disappearance data are incompatible.  But, in reality, the only issue is that there is a background that mimics a possible oscillation signal.

\subsection{Using Markov Chain Methods for Finding Global Fit Parameters}

The process of finding the best fit in a global analysis comes under the umbrella of algorithms called ``optimizers." The most efficient optimizers use a gradient to quickly find the minimum (or maximum) of a function. However, gradient descent is susceptible to being trapped in local minima, and ``homes in'' on the location of the minimum, without exploring the space around it. We are interested not just in the best fit, but also this surrounding region, which characterizes the level of confidence we have in our result. 

Ideally, we would like a method of optimization that moves towards the minimum on average, but otherwise randomly explores the parameter space. The random walk would allow the method to ``tunnel through'' local minima to find other regions of the parameter space. Such a process can be modeled by a Markov Chain \cite{markov} that provides an efficient way to sample the likely regions of the parameter space. 

A simple form of a Markov-chain-based fit proceeds as follows.   At each step, a proposed set of parameters is selected that is a function of the previous parameters in the Markov chain, and the parameters are accepted based only on the set preceding it.   The code makes use of a 
vector of independent parameters $x$.  Each independent parameter is generated and varied according to a function that depends upon the previous step, $x_{old}$, in the minimization chain.    As a simple example, one might use:
\begin{equation}
x = x_{old} + s(R-0.5)(x_{min} - x_{max})~, \label{mcmcstep}
\end{equation}
where $x_{min}$ and $x_{max}$ represent the boundaries on the parameter $x$; $R$ is a random number between 0 and 1, which is varied as one steps from $x_{old}$ to $x$; and $s$ is the ``stepsize.''    The acceptance function for $x$ can, for example, be formed as a Boltzmann distribution:
\begin{equation}
P = min(1,e^{-(\chi^2 -\chi^2_{old})/T})~, \label{mcmcprob}
\end{equation}
where $T$ is the Markov Chain parameter ``temperature.''  The user-defined step-size and temperature control how quickly the Markov Chain diffuses
toward the minimum $\chi^2$ value.   At every step along the chain, each of which
corresponds to a point in the oscillation space described by the vector of independent parameters,
a $\chi^2$ is calculated appropriately for normally or Poisson distributed data. Following a specified number of steps,  the minimum $\chi^2$ is found from the list.  Then, 
\begin{equation}
    \Delta \chi^2(\vec{x}) = \chi^2(\vec{x}) - \chi^2_{\text{min}},
\end{equation}
is calculated for each saved $\chi^2$. These $\Delta \chi^2$ values are used to draw the confidence intervals in plots. All points that satisfy
\begin{equation}
    \Delta \chi^2 < \text{CDF}_{\chi^2}^{-1}(k, p),
\end{equation}
are drawn inside the interval with probability $p$. 
The $\text{CDF}_{\chi^2}^{-1}$ is the inverse $\chi^2$ cumulative distribution function (CDF) and $k$ is the number of degrees of freedom with 
$k=2$ for a two-dimensional plot.  

The simple Markov chain algorithm described above has been used in many of the early fits to short baseline data \cite{sorel,karagiorgi,ignarra}. However, in the intervening years, much progress has been made in the field of Markov chain based optimizers. In particular, algorithms that make efficient use of parallel computing resources have been developed. 

Often called Markov chain Monte-Carlo (MCMC) methods, these algorithms make probabilistic samples of the parameter space, allowing for Bayesian analysis. An example is to use fitting software with an affine invariant parallel tempering MCMC based on the Emcee fitting package \cite{emcee}. Details of this approach are described in Ref.~\cite{SBL2016}.   This approach runs many Markov chains in parallel, and each chain shares information about the parameters space with the other chains. Thus, the algorithms can make more efficient proposals, as it adapts to the shape of the $\chi^2$ surface. 

Drawing the confidence intervals from the  $\Delta\chi^2$ statistic, assumes that the statistic is correctly $\chi^2$ distributed. As discussed above, 
this may not be true, and so the 
Feldman-Cousins technique \cite{Feldman_Cousins} was introduced to allow for meaningful confidence intervals in these conditions. However, the method, which involves throwing many fake experiments for each parameter set in the fit,  is far too computationally expensive to be practical to use in many global fits.   Another way to avoid the issue is to use the log likelihood instead of the $\chi^2$ statistic to determine the probability. This can be done using Bayesian credible intervals, which can be found using the MCMC, as is discussed in Ref.~\cite{SBL2016}.   

\subsection{Global Fit Results}

This section summarizes the status of global fits that make use of the experiments discussed in Sec.~\ref{currentresults}.   The discussion of specific solutions risks becoming dated quickly, and so here we will be brief and take a ``big-picture'' approach to the results.  

\begin{figure}[t]
\vspace{+0.1in}
\centerline{\includegraphics[angle=0, width=1.0\textwidth]{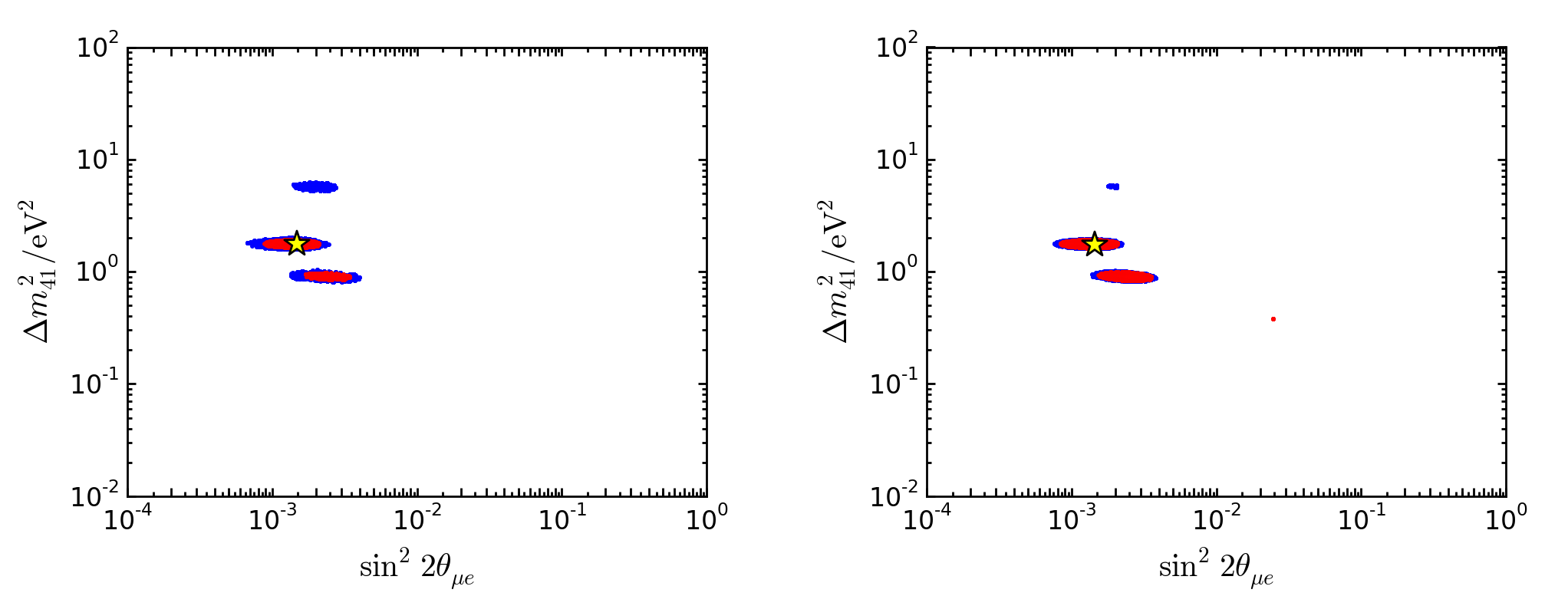}}
\vspace{-0.1in}
\caption{Left:  Global fit to short baseline data, showing the frequentist confidence regions.  Right: Global fit, showing Bayesian credible intervals Ref.~\cite{SBL2016}.} 
\label{globalsbl}
\vspace{-0.1in}
\end{figure}

When the data of Table~\ref{tab:explist} is combined into a global fit, one obtains the result shown in Fig.~\ref{globalsbl} (left).   In this case, three allowed regions are present.    These allowed regions are driving the present interest in sterile neutrino searches.     The allowed regions appear at the overlap of the $\nu_e$ disappearance and $\nu_\mu \rightarrow \nu_e$ appearance signals.  This contour is modified by those experiments with limits, especially the $\nu_\mu$ disappearance limit from SciBooNE/MiniBooNE, which is what is giving the allowed regions the distinct ``three island'' shape.   If one compares the allowed regions of in Fig.~\ref{globalsbl} (left) to the strongly varying oscillations of the SciBooNE/MiniBooNE result shown in Fig.~\ref{fig:LSNDMBSB} (right), one sees that the solutions lie between the strongly varying SciBooNE/MiniBooNE contours.      

This analysis draws confidence intervals from the $\Delta\chi^2$ statistic, assuming that this statistic is correctly described by a $\chi^2$ distribution with $dof$ equal to the number of fit parameters. In the previous section, we pointed out that this assumption may be incorrect for a number of reasons.  A comparison to the result from Bayesian credible intervals, Fig.~\ref{globalsbl} (right), found using the MCMC method previously discussed and described in Ref.~\cite{SBL2016},  shows that the results for the two global fits are very similar.  Thus, this check indicates that the $\Delta\chi^2$ distribution assumption must be relatively good.

\begin{figure}[t]
\vspace{+0.1in}
\centerline{\includegraphics[angle=0, width=0.75\textwidth]{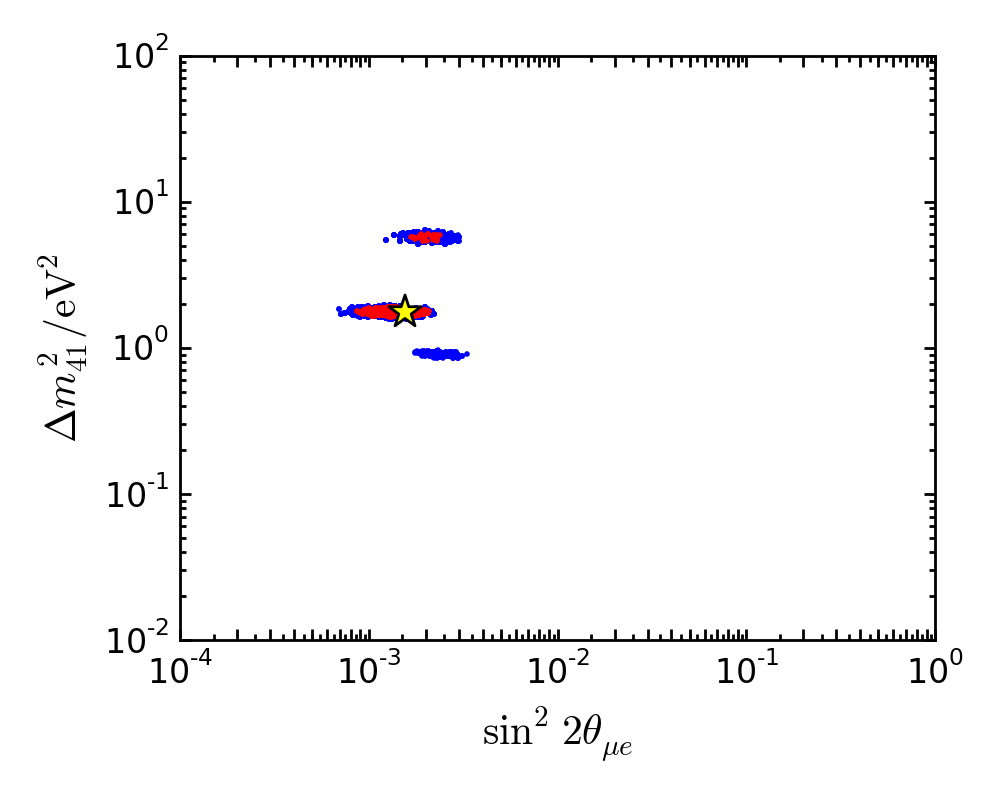}}
\vspace{-0.1in}
\caption{Frequentist global fit including the IceCube data, from Ref.~\cite{icefits}.} 
\label{iceglob}
\vspace{-0.1in}
\end{figure}

Including the IceCube result is quite time consuming to include because of the need to repeatedly propagate the atmospheric flux through the matter profile of the earth, varying the fit parameters with each iteration.   Thus, these data were included in the fit by first fitting the short baseline data sets, determining the parameter range of interest, and then addressing IceCube for only the viable models.  Details of the method are described in Ref. \cite{icefits}.   The result of this approach leads to Fig.~\ref{iceglob}.   The IceCube result substantially weakens the $\sim 1$ eV$^2$ allowed region.   With this said, it should be noted that including IceCube data in the fits assumes the simplest 3+1 oscillation model with no additional Beyond-Standard-Model physics.  Introducing new physics that would obscure or inhibit the sterile-neutrino related matter effects, would weaken the IceCube result.    Thus, the IceCube data requires some assumptions beyond the simplest 3+1 model.

The results of the global fits, with and without IceCube, are summarized in Table~\ref{tab:bfpointsIce} results.   As per the discussion above, the interesting parameters to compare are the $\chi^2$ for the global best fit point compared to the  no oscillations (``null'') value.    One can see that the difference is on the order of 50 compared to the 3 or 4 value for the $dof$'s.    Thus the $\Delta \chi^2$ method is indicating a strong preference for a fit that includes a sterile neutrino.    Note that the difference in the null and best fit $\chi^2$ for the IceCube only fit is small.  This is the hallmark of the case where the best fit point is well outside of the region addressed by the data set.  Although the IceCube result has not changed the best-fit point, it does guide future experiments to concentrate on regions that are generally have $\Delta m^2 > 1$ eV$^2$.

\begin{table*}[t]
\begin{center}
\begin{tabular}{|c|cccc|cccc|}\hline
3+1 & $\Delta m^2_{41}$ & $|U_{e 4}|$ & $|U_{\mu 4}|$ & $|U_{\tau 4}|$ & $N_{bins}$ & $\chi^2_{\mathrm{min}}$ & $\chi^2_{\mathrm{null}}$ & $\Delta \chi^2$ (dof)  \ \\  
\hline \hline
SBL  & 1.75 & 0.163 & 0.117 & - & 315 & 306.81 & 359.15  &52.34 (3)  \\  \hline
SBL+IC & 1.75 & 0.164  & 0.119 & 0.00  & 524 & 518.23 & 568.84 & 50.61 (4)\\ \hline
IC & 5.62 & - & 0.314 & - & 209 & 207.11 & 209.69 & 2.58 (2) \\ \hline
\end{tabular}
\end{center}

\caption{The oscillation parameter best-fit points for $3+1$ for the combined short baseline (SBL) and IceCube (IC) data sets compared to SBL alone, as reported in Ref.~\cite{icefits}.   Units of $\Delta m^2$ are eV${}^2$.  
\label{tab:bfpointsIce}}
\end{table*}

At the time of writing, 3+2 and 3+3 fits have been performed for the short-baseline data sets only.   Including IceCube in the fits is even more computationally time-consuming than for 3+1 due to the additional parameters.     However, since the IceCube result did not strongly affect the best fit for 3+1, it is likely that it will also not strongly affect the 3+2 and 3+3 results. It is found that the short-baseline-fit result for 3+2 is nearly the same as 3+1. The 3+2 fit has a $\Delta \chi^2$[null-min] (dof) of 56.99 (7).  Thus adding additional parameters does not affect the conclusions.    It is for this reason, and because the fits are quicker and simpler, that, at present, usually only 3+1 fits are discussed in conferences.

\section{Future Experiments and Requirements}

The field of sterile neutrino experiments is rapidly developing.   Follow-up experiments that build on the techniques of the experiments listed in Table~\ref{tab:explist} are proliferating rapidly.   Even more exciting, new and innovative technologies and techniques are being developed to explore electron flavor disappearance (see, for example, IsoDAR \cite{isoDAR}), muon flavor disappearance (see, for example, KPIPE \cite{KPIPE}), and muon-to-electron appearance experiments (see, for example, MicroBooNE \cite{uB}) are underway.  
A review of all of the planned experiments is beyond the scope of this chapter.  Instead, for a review, we refer the reader to articles that are routinely appearing on the arXiv. 
Instead, here, we approach this discussion of future experiments from an educational point-of-view.    From our summary of the existing results, above, one can identify requirements for future experiments to address the current anomalies and unexplored oscillation regions.     We look at these requirements in detail below.

The current muon to electron flavor ($\nu_\mu \rightarrow \nu_e$) appearance signals are the strongest indications for oscillations in the $\Delta m^2$ around 1 eV$^2$ region.   However, as discussed above, there are tensions between MiniBooNE neutrino data and antineutrino results from MiniBooNE and LSND.   Thus, it is essential that the MiniBooNE neutrino result be checked with improved and higher statistics measurements.  Here, a new technology,  liquid argon time projection detectors (LArTPCs) \cite{LArReview} have the promise of large data samples with improved background rejection and measurement resolution. 
These systems are large volumes of liquid argon with two
subdetectors:  a time projection chamber (TPC) for tracking, and a light
collection system.    The MicroBooNE experiment, described in Ref.~\cite{uB}, is an example.     LArTPCs have the capability to discriminate electron showers from background $\gamma$ showers using the $dE/dx$ at the start of the shower.   This type of experiment can, therefore, address the potential of a background contamination in the MiniBooNE signal that might arise from electron/$\gamma$ confusion in a Cherenkov detector.  MicroBooNE has just begun an experimental run at Fermilab, on the same beamline and near in location to MiniBooNE.    In the future, MicroBooNE is to be combined with a near LArTPC detector (SBND) and a large far LArTPC detector (ICARUS) to form the Short Baseline Neutrino (SBN)\cite{SBN} program at Fermilab.  This program is being designed to have good sensitivity in the regions associated with the current signals by comparing the rates of appearance in the three detectors.  This comparison will help constrain systematic uncertainties.  The comparison also will have some ability to map out the expected oscillatory behavior of sterile oscillations, in that it will measure the observed neutrino rate versus energy at three points.   

\begin{figure}[t]
\begin{center}
\includegraphics[scale=.5]{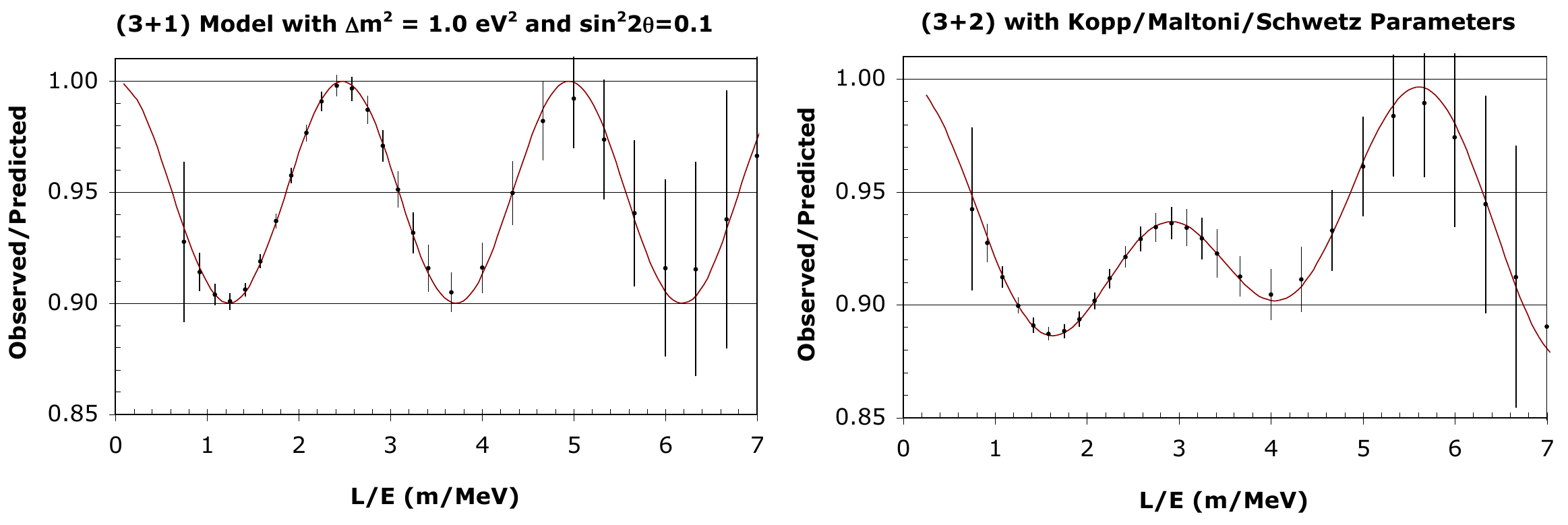}
\end{center}
\vspace{-.5cm}
\caption{\footnotesize The $L/E$ dependence of sample data sets for 5 years of running for 3+1 (left) and 3+2
  (right) oscillation scenarios.   The solid curve is the oscillation probability with no smearing in the reconstructed position and energy
  and the data points with error bars are from simulated events including smearing.
\label{waves}}
\end{figure}

Improved searches for electron neutrino disappearance are also a prime area for improved experiments.  Several radioactive source experiments are being considered using existing large detectors \cite{new_rad_source}.   But the problem with these experiments is that the very hot radioactive sources are short-lived, and so the experiment can only run for a modest period of time.   As a result, the sensitivity is  typically at the level to only just cover the ``reactor anomaly'' region.  Many new very short baseline reactor oscillation experiments \cite{new_reactor} are also being planned with some prototypes underway.  These experiments need to have setups within ~10 m from the reactor core and so have to contend with high background rates from neutrons and gammas, that can have confusing position dependence, as well as the large source size of the reactor core, which smears the signal.   The reactor flux is also a combination of many decays.    It is possible that some decay sources are isolated to specific regions of the reactor, because of the arrangement of fuel.  The IsoDAR experiment \cite{isoDAR} avoids many of these issues by creating higher energy (8 MeV) electron antineutrinos from $^8$Li beta decay.    This has the advantages of a very intense radioactive source experiments, but can run for long periods because this is a ``driven source.''  By this we mean that the $^8$Li is constantly produced from $^7$Li using a 60 MeV, 10 mA proton beam from a cyclotron that the collaboration is developing.  This $^8$Li source provides an isotropic source of antineutrinos that would be placed near a large scintillator neutrino detector such as KamLAND or JUNO \cite{isokamjuno}.  For these setups, oscillation signals would be detected by observing the oscillatory behavior within the detector, thus, minimizing systematic uncertainties and providing a definitive signal of oscillations.  The oscillatory pattern even has the capability to separate 3+1 from 3+2 oscillation models, as is shown in Fig.~\ref{waves}.    

In general, ability to reconstruct the oscillation wave in detail, as is shown in Fig.~\ref{waves}, should be a goal for future neutrino experiments.  As discussed above, the actual sterile neutrino model that describes neutrino oscillations may be much more complicated than the 3+1 world-view.     However, all of these models will predict oscillation waves.   Thus, patterns like these are the ``smoking gun'' for neutrino oscillations involving sterile neutrinos.

For muon neutrino disappearance, new experiments must improve on the past experiments with better statistics and reduced systematic uncertainties.  Improved results from IceCube are expected with somewhat better sensitivity to high $\Delta m^2$ and mixing angle.  The multi-detector SBN program also has good sensitivity to muon neutrino disappearance by comparing the detected rate versus energy in the three detectors and can make measurements with both $\nu_\mu$ and $\bar\nu_\mu$ beam settings.    However, the KPIPE experiment \cite{KPIPE} is, at present,  the one example of a $\nu_\mu$ disappearance experiment that can observe and use the oscillatory behavior of oscillations to detect a disappearance signal.  The neutrino source is from the decay of stopped kaons produced in the J-PARC MLF spallation target, which produces mono-energetic 236 MeV muon neutrinos.    Note that a mono-energetic neutrino source that has high enough energy to produce charge-current $\nu_\mu$ interactions is unique among experiments.   In this case, the $L/E$ dependence reduced to just $L$ dependence, since $E$ is a constant.  A 120 m long by 3 m diameter cylindrical liquid-scintillator detector is proposed to be used to detect the neutrino rate as a function of distance.  Like IsoDAR, this allows for a search for the oscillatory behavior of the signal, with coverage beyond the other proposed experiments. 

These examples give you a taste of the interesting new ideas that are now under discussion.
Exploring the possibility that light sterile neutrinos exist is one of the current major goals of particle physics.   If sterile neutrinos are established, this will be revolutionary for the field.  Previous measurements are a possible guide, particularly when considered within the context of global fits,  and have led to an extensive list of new experiments and programs.  These experiments are in the small to mid-scale range  and, thus, are real possibilities to be supported and initiated over the next decade.   That makes this a great time to be interested in the physics of this thriving field of sterile neutrino searches.

\section*{Acknowledgements}

The authors thank the members of their research groups for discussion, and especially thank Carlos Arg\"uelles, Leslie Camilleri, Gabriel Collin, J. I. Crespo-Anad\'{o}n, Alejandro Diaz, Ben J.P. Jones, Boris Kayser, Blake Watson and Lauren Yates for their suggestions on the text.   JMC is supported by NSF grant 1505855 and 1505858.  MHS is supported by NSF grant 1404209.

\end{document}